\documentclass[fleqn,usenatbib]{mnras}

\usepackage{newtxtext,newtxmath}
\usepackage[T1]{fontenc}

\DeclareRobustCommand{\VAN}[3]{#2}
\let\VANthebibliography\thebibliography
\def\thebibliography{\DeclareRobustCommand{\VAN}[3]{##3}\VANthebibliography}

\usepackage{graphicx}	
\usepackage{amsmath}	
\usepackage{lipsum}  
\usepackage[dvipsnames]{xcolor}
\usepackage{physics}

\definecolor{sim_red}{HTML}{EF476F}
\definecolor{sim_yellow}{HTML}{FFD166}
\definecolor{sim_green}{HTML}{06D6A0}
\definecolor{sim_blue}{HTML}{118AB2}
\definecolor{sim_purple}{HTML}{9C12E4}

\usepackage{scalerel}
\usepackage{tikz}
\usetikzlibrary{svg.path}

\definecolor{orcidlogocol}{HTML}{A6CE39}
\tikzset{
  orcidlogo/.pic={
    \fill[orcidlogocol] svg{M256,128c0,70.7-57.3,128-128,128C57.3,256,0,198.7,0,128C0,57.3,57.3,0,128,0C198.7,0,256,57.3,256,128z};
    \fill[white] svg{M86.3,186.2H70.9V79.1h15.4v48.4V186.2z}
                 svg{M108.9,79.1h41.6c39.6,0,57,28.3,57,53.6c0,27.5-21.5,53.6-56.8,53.6h-41.8V79.1z M124.3,172.4h24.5c34.9,0,42.9-26.5,42.9-39.7c0-21.5-13.7-39.7-43.7-39.7h-23.7V172.4z}
                 svg{M88.7,56.8c0,5.5-4.5,10.1-10.1,10.1c-5.6,0-10.1-4.6-10.1-10.1c0-5.6,4.5-10.1,10.1-10.1C84.2,46.7,88.7,51.3,88.7,56.8z};
  }
}

\newcommand\orcidicon[1]{\href{https://orcid.org/#1}{\mbox{\scalerel*{
\begin{tikzpicture}[yscale=-1,transform shape]
\pic{orcidlogo};
\end{tikzpicture}
}{|}}}}

\usepackage{hyperref}
\usepackage{ulem}

\title[AGN at Cosmic Dawn]{Supermassive Black Hole Growth in Massive Galaxies at Cosmic Dawn}

\author[Sunseri et al.]{James Sunseri \orcidicon{0000-0003-4274-2662},$^{1}$\thanks{E-mail: jsunseri@princeton.edu (JS)}
Zachary L. Andalman \orcidicon{0000-0001-5064-1269},$^{1}$
and Romain Teyssier \orcidicon{0000-0001-7689-0933}$^{1}$
\\ \\
$^{1}$Department of Astrophysical Sciences, Princeton University, Princeton, NJ 08540, USA \\
}

\date{Accepted XXX. Received YYY; in original form ZZZ}

\pubyear{2025}

\begin{document}
\label{firstpage}
\pagerange{\pageref{firstpage}--\pageref{lastpage}}
\maketitle

\begin{abstract}
Among the emerging excess of massive, bright galaxies at Cosmic Dawn $z \gtrsim 9$ seen by the James Webb Space Telescope, several exhibit spectral features associated with active galactic nuclei (AGN). These AGN candidates suggest that supermassive black holes (SMBHs) grow rapidly in the early Universe. In a series of numerical experiments, we investigate how SMBHs grow within and influence the most massive galaxies at Cosmic Dawn using cosmological hydrodynamic zoom-in simulations run with the adaptive mesh refinement code \textsc{ramses}. Our suite of simulations explore how super-Eddington accretion, seed mass, and the strength of feedback influence SMBH-galaxy co-evolution in the most massive galaxies ($M_\star \gtrsim 10^8 M_\odot$) of the early Universe ($z \sim 15 - 9$). The environment which our numerical experiments reside in is an overdensity that collapses into a $\sim 10^{11} M_\odot$ halo by $z \sim 9$. Within this type of environment we find that SMBH growth is sensitive to stellar feedback which generates a turbulent-multiphase interstellar medium (ISM) that stochastically starves the SMBH. In the absence of AGN feedback, we find that the SMBH is starved $\sim 50\%$ of the time after the onset of star formation in the galaxy. SMBH growth can become self-regulated by AGN feedback if the SMBH becomes massive enough, either by accretion or seeding, for its feedback to dominate the surrounding nuclear region. We find no evidence of galaxy-scale, AGN-driven quenching in the star formation rate (SFR) across all simulations in our suite. 
\end{abstract}

\begin{keywords}
galaxies: supermassive black holes -- cosmology: early Universe -- galaxies: formation
\end{keywords}



\section{Introduction}
\label{sec:intro}
Since the start of the decade, the James Webb Space Telescope (\textit{JWST}) has discovered a plethora of massive galaxies at Cosmic Dawn ($z \gtrsim 9$) with high stellar masses $M_\star \gtrsim 10^8~\rm M_\odot$ and/or high star formation rates (SFRs) $\dot{M}_{\star} \gtrsim 1~\rm M_\odot/yr$ \citep{finkelstein_ceers_2023, labbe_population_2023, mason_brightest_2023, carniani_spectroscopic_2024} that are inconsistent with expectations from standard models of galaxy formation \citep{boylan-kolchin_stress_2023}. The UV luminosity function (UVLF) inferred from these Massive Dawn Galaxies (MDGs) suggests that $\sim 30\%$ of the baryonic matter accreting onto the host halo must be converted into stars \citep{shen_impact_2023}. This star-formation efficiency (SFE) is high relative to the typical SFEs $\lesssim 5\%$ in the local Universe inferred from abundance matching \citep{rodriguez-puebla_constraining_2017, moster_emerge_2018, behroozi_universemachine_2019}.  
 
\citet{dekel_efficient_2023} suggest that star formation may be intrinsically more efficient in MDGs due to high gas densities which suppress stellar feedback. Simple analytic estimates suggest that gas number densities can reach up to $10^3~{\rm cm^{-3}}$ in galaxies hosted by haloes with masses $\sim 10^{11}~{\rm M_\odot}$ at redshift $z=9$ \citep{dekel_efficient_2023,andalman_origin_2025}. These densities are seldom found in the local Universe aside from the nuclear regions of starburst galaxies.

At Cosmic Dawn, cold gas rapidly accretes onto MDGs through cosmic filaments, seeding strong turbulence \citep{dekel_cold_2009, ginzburg_evolution_2022}. This picture is consistent with the high velocity dispersions $\approx 30 - 70$ km/s observed in high redshift galaxies \citep{de_graaff_ionised_2024}. 
The effect of strong turbulence on star formation and stellar feedback is not captured by standard star formation models used in big box cosmological simulations like \textsc{Eagle} \citep{schaye_eagle_2015, crain_eagle_2015}, \textsc{Flamingo} \citep{schaye_2023_flamingo}, \textsc{Illustris-TNG} \citep{naiman_first_2018, nelson_first_2018, marinacci_first_2018, springel_first_2018, pillepich_simulating_2018}, \textsc{Sphinx} \citep{rosdahl_sphinx_2018}, \textsc{Simba} \cite{dave_simba_2019}, \textsc{Cosmic Dawn II} \citep{ocvirk_cosmic_2020}, and \textsc{Thesan} \citep{kannan_introducing_2022}. 
It is currently not computationally feasible to simulate the large volumes $\gtrsim (100 \;h^{-1} \rm cMpc)^3$ required to sample massive halos while simultaneously resolving the turbulent structure of the interstellar medium (ISM) on scales $\sim 10 \; \rm pc$.
For this reason, these simulations use a simple and computationally-efficient star formation prescription based on the Kennicut-Schmidt Law \citep{kennicutt_star_1998, kennicutt_global_1998}, tuned to match observations in the local Universe \citep[e.g.][]{hopkins_galaxies_2014}.

Cosmological zoom-in simulations informed by big box dark-matter-only simulations offer a solution to this resolution problem. The dark-matter-only simulations are used to preselect halos of interest which are then re-simulated in a smaller box with higher resolution and full hydrodynamics. This technique is used in several recent simulations including \textsc{Renaissance} \citep{oshea_2015_renaissance}, \textsc{FirstLight} \citep{ceverino_2017_firstlight}, \textsc{Fire-2} \citep{ma_2018_fire2}, \textsc{Flares} \citep{lovell_2021_flares, vijayan_2021_flares}, \textsc{Serra} \citep{pallottini_2022_serra}, and \textsc{Thesan-Zoom} \citep{kannan_2025_thesanzoom}. While these simulations were originally designed to forward model observations and population properties like the UVLF, their use has been extended to better understand efficient star formation in the early Universe \citep{bassini_2023_sfe, ceverino_2024_sfe, shen_2025_sfe}. These studies show systematically higher star formation efficiencies in massive galaxies at high redshifts, where there are high gas density conditions.

\cite{andalman_origin_2025} ran cosmological zoom-in simulations using the adaptive mesh refinement code \textsc{ramses} \citep{teyssier_cosmological_2002} and a physically-motivated multi-freefall model \citep{federrath_star_2012} to predict the SFR from a subgrid model of turbulence. With their physically motivated subgrid model of star-formation and ability to capture large-scale accretion from cosmic filaments ($L = 100 h^{-1} \; \rm cMpc$) while resolving the turbulent structure of the interstellar medium (ISM) within the galaxy on scales $\Delta x_{\rm min} \sim 10 \; \rm pc$, they produced a MDG-like galaxy with a high stellar mass $M_\star \sim 10^9 \; \rm M_\odot$, high SFR $\sim 50 \; \rm M_\odot /yr$, and a high local SFE $\gtrsim 10 \%$ by $z \sim 9$. While these simulations were successful in producing MDG-like galaxies, they neglected the role of the central supermassive black hole (SMBH).

In the local Universe, SMBHs are ubiquitously found at the centres of galaxies and co-evolve with their host \citep[e.g.][]{ciotti_radiative_2007, fabian_observational_2012, kormendy_coevolution_2013, heckman_coevolution_2014, reines_relations_2015, greene_role_2020, bennert_local_2021}. Central SMBHs in active galactic nuclei (AGN) produce feedback in the form of winds and jets which heat and ionize the surrounding gas, thus regulating star formation \citep{silk_quasars_1998, di_matteo_energy_2005, mountrichas_link_2023, goubert_role_2024, bluck_galaxy_2024}. The mass of the SMBH is tightly correlated with various host galaxy properties (e.g. stellar mass, velocity dispersion of the bulge, and bolometric luminosity) which likely evolve with redshift \citep{kauffmann_host_2003, merloni_cosmic_2009, jahnke_massive_2009,bennert_relation_2011, shen_sloan_2015, suh_no_2020, ding_mass_2020,li_assembly_2023, farrah_observational_2023, graham_appreciating_2023}. 

Some MDGs exhibit X-ray emission, high ionization lines, and/or broad line features indicative of AGN \citep{kocevski_hidden_2023, harikane_jwstnirspec_2023,  larson_ceers_2023, ubler_ga-nifs_2023, ubler_ga-nifs_2024, ubler_ga-nifs_2024-1,  maiolino_small_2024, maiolino_jades_2024-1, maiolino_jades_2024, matthee_little_2024, greene_uncover_2024, juodzbalis_jades_2024, juodzbalis_dormant_2024, scholtz_jades_2025, perna_ga-nifs_2025}. Local virial relations have been used to estimate SMBH masses for a subset of these galaxies \citep{kocevski_hidden_2023, ubler_ga-nifs_2023, harikane_jwstnirspec_2023, maiolino_black_2025, taylor_capers-lrd-z9_2025, tripodi_red_2024, juodzbalis_dormant_2024}, but the reliability of those mass estimates is debated \citep{juodzbalis_jades_2025, naidu_black_2025, rusakov_jwsts_2025}. Paired with estimates of host galaxy stellar mass \citep{harikane_jwstnirspec_2023, maiolino_jades_2024, juodzbalis_jades_2025}, SMBH mass estimates indicate that several SMBHs at Cosmic Dawn are overmassive relative to local $M_{\rm BH}-M_\star$ relations by at least 1-2 dex \citep{ubler_ga-nifs_2023, ubler_ga-nifs_2024, harikane_jwstnirspec_2023, kokorev_uncover_2023, carnall_massive_2023, maiolino_jades_2024, pacucci_jwst_2023,  maiolino_small_2024, furtak_high_2024, juodzbalis_dormant_2024, natarajan_first_2024, andika_tracing_2024}, although some of the discrepancy may arise from selection biases \citep{ananna_x-ray_2024, lupi_size_2024}. Regardless, these empirical estimates require a mechanism to produce overmassive SMBHs at Cosmic Dawn. The prominent theoretical proposals invoke either the formation of heavy seeds $\gtrsim 10^4~M_\odot$ or super-Eddington accretion. 

SMBH seed formation scenarios are typically separated into 3 groups segmented by resulting seed mass: light seeds $\sim 10^2~M_\odot$, intermediate seeds $10^{\sim 2-3}~M_\odot$, and heavy seeds $\sim 10^{4-6}~M_\odot$. Light seeds are produced by black hole remnants of Population III stars \citep[e.g.][]{madau_massive_2001, schaerer_properties_2002}. Intermediate seeds are likely produced by runaway mergers \citep[e.g.][]{portegies_zwart_formation_2004, freitag_stellar_2006, lupi_constraining_2014, reinoso_collisions_2018, arca_sedda_dragon-ii_2024, gaete_supermassive_2024, rantala_rapid_2025, dekel_2025_ffb_imbh}. Heavy seeds are produced by the direct collapse of gas in halos of size $\sim 10^{7-8} \; \rm M_\odot$ \citep[e.g.][]{oh_second-generation_2002, begelman_formation_2006, natarajan_2011_bh_seed_review, latif_black_2013, latif_formation_2014, latif_formation_2016, wise_formation_2019, basu_mass_2019, narayan_black_2023, shen_massive_2025}. Light and intermediate seed scenarios require super-Eddington accretion to be consistent with the massive SMBHs inferred from recent \textit{JWST} observations \citep{volonteri_rapid_2005, madau_super-critical_2014, volonteri_evolution_2016, pezzulli_super-eddington_2016, inayoshi_assembly_2020, schneider_are_2023, massonneau_how_2023, sassano_super-critical_2023, bennett_growth_2024, pacucci_mildly_2024,lupi_sustained_2024, lupi_size_2024, trinca_episodic_2024,king_joining_2025, husko_effects_2025, quadri_super-eddington_2025, sanati_rapid_2025}. Super-Eddington accretion is already well studied in tidal disruption events, ultraluminous X-ray sources, and quasars \citep[see][for a review]{jiang_numerical_2024}. Several studies have explored super-Eddington accretion onto SMBHs in cosmological environments \citep{di_matteo_origin_2017, regan_super-eddington_2019, angles-alcazar_cosmological_2021, massonneau_how_2023, rennehan_manhattan_2024, lupi_sustained_2024, gordon_conditions_2025, husko_effects_2025, quadri_super-eddington_2025, sanati_rapid_2025}. These works have shown that, while super-Eddington accretion is a viable option for rapid SMBH growth, the impact of AGN and stellar feedback on this growth is still not fully understood across all possible environments.

In this work, we explore in detail, how SMBHs grow in and influence the most massive galaxies at Cosmic Dawn. We do so by introducing SMBHs into the precursory simulations of \cite{andalman_origin_2025} originally designed to replicate the MDGs commonly seen by \textit{JWST}. We run a suite of 11 numerical experiments exploring how AGN feedback, SMBH seed mass, and maximal accretion rates impact the growth of the SMBH and the co-evolution of the SMBH with the host galaxy. In Section~\ref{sec:simulations} we introduce the simulations used in our analysis and in Section~\ref{sec:results} we discuss how the SMBH grows and regulates its own growth via AGN feedback (self-regulates) across all of the simulations. We also discuss the impact of the SMBH on the host galaxy and provide insights on growth from relevant timescales in the simulations. Lastly, in Section~\ref{sec:conclusions} we summarize our findings from this work.

\section{Simulations}
\label{sec:simulations}
\begin{figure*}
    \centering
\includegraphics[width=2\columnwidth]{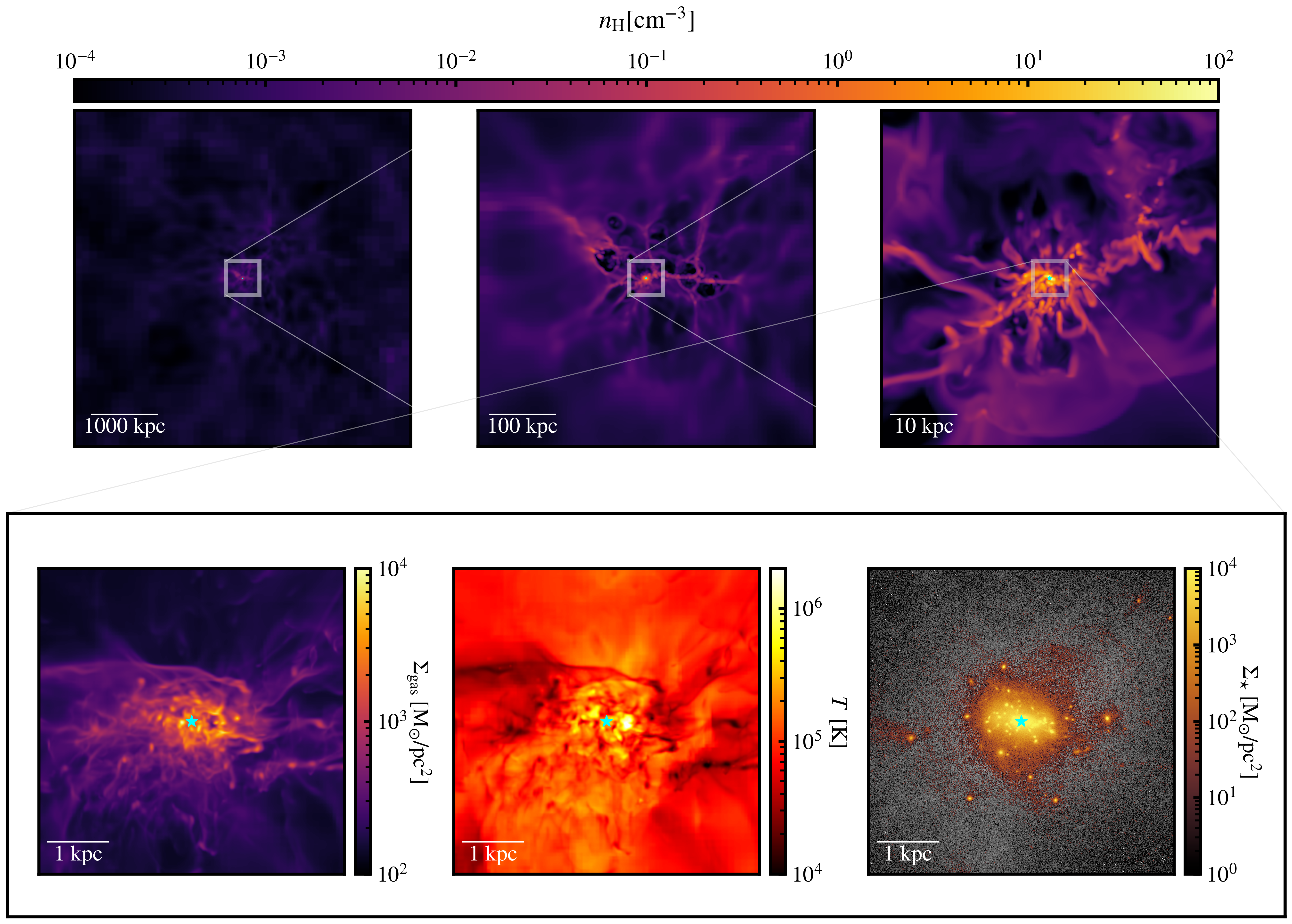}
    \caption{This figure depicts our zoom-in simulation of a massive Cosmic Dawn galaxy at $z = 9$ with a SMBH at the centre regulated by AGN feedback. The top row shows consecutive zooms into a slice of the gas density depicting gas filaments which feed onto the host galaxy. In the bottom row we show projected quantities: the gas surface density, temperature (weighted by gas density), and the stellar surface density superimposed on the dark matter surface density. The cyan star is the SMBH at the centre of the galaxy. The galaxy resembles a clumpy thick disk of stellar clusters.}
    \label{fig:zoom-in}
\end{figure*}

In this work, we use the adaptive mesh refinement (AMR) code \textsc{ramses} \citep{teyssier_cosmological_2002}. \textsc{ramses} simulates the interactions between gas, star clusters, dark matter, and sink (SMBH) particles in a cosmological context. In this section, we describe our simulations of MDGs with central SMBHs as seen in Fig. \ref{fig:zoom-in}.

\subsection{RAMSES}
\label{subsec: RAMSES}

We build off the simulation setup in \citet{andalman_origin_2025}, which we recap here for clarity. Throughout this work, we adopt a standard $\Lambda$CDM cosmology with parameters $h = 0.7$, $\Omega_{\rm m,0} = 0.276$, $\Omega_{\rm b,0} = 0.049$, $\Omega_{\Lambda, 0} = 0.724$. \textsc{ramses} evolves gas on an adaptively-refined numerical grid. Simultaneously, \textsc{ramses} evolves star cluster, dark matter, and sink particles with a Lagrangian approach. Gas and dark matter are coupled by the Poisson equation. Gas and star cluster particles are coupled by star formation and stellar feedback. Sink particles are evolved using a direct summation approach \citep{bleuler_towards_2014}. 

\textsc{ramses} evolves the hydrodynamical variables (density $\rho$, pressure $P$, velocity $\mathbf{v}$, and metallicity $Z$) on the grid by solving the Euler equations with an unsplit second-order Godunov method and an ideal gas equation of state with $\gamma = 5/3$. \textsc{ramses} sets the timestep adaptively by the stability criteria described in \citet{teyssier_cosmological_2002}. In practice, the typical timestep is $\sim 500~{\rm yr}$. Each dark matter or star cluster particle is prescribed a unique identification number (ID), position $\mathbf{x}$, and velocity $\mathbf{v}$. Star cluster particles also carry additional metadata: mass $m_\star$, metallicity $Z$, and time of birth $t_{\rm birth}$. 

We generate initial conditions using \textsc{music} \citep{hahn_multi-scale_2011}. The initial conditions are chosen such that the galaxy of interest will form at the centre of our simulation domain: a periodic box of comoving volume $(100 h^{-1} \; \rm cMpc)^{3}$. This is done by first generating initial conditions at $z=100$ with \textsc{music} for a fixed low-resolution ($\Delta x \approx 0.2 h^{-1} \; \rm cMpc$, $m_{\rm dm} = 1.72 \times 10^6 \; \rm M_\odot$)\; dark matter only simulation such that a candidate halo with mass $M_{\rm halo} \simeq 10^{11} \; \rm M_\odot$ forms within the domain by $z=9$. From this precursory simulation, a new set of initial conditions are generated with \textsc{music} including baryons and dark matter, translated so that the candidate halo forms near the centre of the box by $z=9$.

Throughout the simulation the grid mesh refines adaptively, such that the width of a grid cell can be written as $\Delta x = L / 2^\ell$ where $\ell$ is the refinement level. As the mesh refines the mass resolution of dark matter particles is given by $m_{\rm dm} = \Omega_{\rm m,0} \rho_{\rm crit,0} \left( L / 2^\ell \right)^3$. Within the zoom-in region of interest, the minimum mass resolution of the dark matter particles is $m_{\rm dm, min} \simeq 2.48 \times 10^4 \; \rm M_\odot$ ($\ell = 14$). This minimum dark matter mass resolution corresponds to a baryon mass resolution $m_{\rm b, min} = f_{\rm b} m_{\rm dm, min}$ where $f_{\rm b} = \Omega_{\rm b,0} / \Omega_{\rm m, 0} \simeq 0.17$. Outside the zoom region of interest, the mass resolution degrades down to a minimum refinement level of $\ell = 7$. In order for a grid cell to refine within the simulation, the advected refinement mask $\texttt{M}$ must be greater than 0.1 and either the mass of dark matter within the cell must exceed 8 $m_{\rm dm}$, the baryonic mass in the cell exceeds $8m_{\rm b}$, or more than 5 Jeans lengths are within the cell. These conditions ensure that cells at the highest refinement level will have the ability to form stars. 

As the simulation runs, the domain expands with the Hubble flow. Each time the physical box size doubles we increment the maximum refinement level $\ell_{\rm max}$. We stop refining beyond $\ell_{\rm max} = 20$ occurring when the scale factor $a \geq 0.05$. This refinement is done to maintain a physical grid resolution of $\Delta x_{\rm min} \simeq 10 \; \rm pc$ in the zoom-in region over the entire time evolution of the simulation.

All simulations in this work make use of the Stellar cluster at Princeton University. Each simulation requires $\sim 130,000$ CPU hours running on 2.9 GHz Intel Cascade Lake CPUs. 

\subsection{Simulating Massive Cosmic Dawn Galaxies}
\label{subsec: simulating MDGs}

The simulations of \cite{andalman_origin_2025} are designed to simulate the rare MDGs observed by \textit{JWST} which commonly have stellar masses $M_\star \geq 10^8 \; \rm M_\odot$ and high SFRs $\dot{M}_\star \gtrsim 1 \; \rm M_\odot / yr$ \citep{finkelstein_ceers_2023, labbe_population_2023, mason_brightest_2023, carniani_spectroscopic_2024}. These galaxies exceed expectations of the UV luminosity function predicted by standard galaxy formation models \citep{boylan-kolchin_stress_2023, shen_impact_2023}, possibly requiring star formation to be $\sim 5 $ times more efficient than what is seen in the local Universe. We use the fiducial simulation of \cite{andalman_origin_2025} as a starting point to investigate the role of SMBH growth and feedback in MDGs. 

\subsubsection{Star Formation} These simulations avoid extrapolations of empirical relations of star formation from the local Universe by using a physically-motivated turbulence-based subgrid multi-freefall star formation model. This allows for a variable local SFE which can fluctuate across the galaxy depending on subgrid turbulence as opposed to a fixed global SFE across the galaxy. 

We evolve the turbulent kinetic energy as a passive scalar using the Large Eddy Simulation (LES) method \citep{kretschmer_forming_2020} with compressive turbulence forcing \citep{federrath_comparing_2010}. In the multi-freefall star formation model, the local SFE per free-fall time is calculated using the turbulent kinetic energy assuming that the density fluctuations follow a log-normal distribution as expected in a supersonic turbulent medium \citep{vazquez-semadeni_hierarchical_1994, kritsuk_statistics_2007}. In each cell, we sample the number of star cluster particles per timestep from a Poisson distribution to match the desired SFR. Each star cluster particle has a mass $M_{\rm cl} = m_{\rm b, min} = 4400~M_\odot$.

We avoid modelling Pop III stars by enriching the ISM to $10^{-3} Z_\odot$ in the initial conditions, surpassing the maximal metallicity which Pop III stars can form from due to fine-structure Carbon and Oxygen line cooling \citep{bromm_formation_2003}. We add cooling terms to the energy equation based on the local metallicity, assuming equilibrium chemistry of Hydrogen and Helium \citep{sutherland_cooling_1993, katz_cosmological_1996}. We assume cosmological abundances of Hydrogen $X = 0.76$ and Helium $Y = 0.24$. We include additional heating for diffuse gas $n_{\rm H} \leq 10^{-2} \; \rm cm^{-3}$ from a time-dependent uniform extragalactic UV background \citep{haardt_radiative_1996, aubert_reionization_2010}. We floor the temperature at the cosmic microwave background (CMB) temperature at $z=9$ given by $T_{\rm CMB} \simeq 2.275~{\rm K}\times (1+z) = 27.25~{\rm K}$.

\subsubsection{Stellar Feedback} The prescription for stellar feedback can be broken up into two categories, early and late type feedback. In both cases, stellar feedback is due to massive $\gtrsim 8~M_\odot$ stars. Early type feedback has multiple sources. Here, we only model thermal pressure associated with photoionization. We note that the efficiency of photoionization feedback is sensitive to the star particle mass $M_{\rm cl}$, we adopt the fiducial value $M_{\rm cl} = 4400 \; \rm M_\odot$ chosen by \cite{andalman_origin_2025}. For further discussion on the impact of this choice, see Section 2.5 of \citet{andalman_origin_2025}. Late type feedback is due to Type II supernovae (SNe), which occur $3-20~{\rm Myr}$ after the birth of the star \citep{leitherer_starburst99_1999, kimm_towards_2015}.

After $20~{\rm Myr}$, all of the massive stars in a cluster will have exploded. Therefore, we only model early feedback from star cluster particles younger than $20~{\rm Myr}$. We set the cell temperature to $20000~{\rm K}$ in cells containing young star cluster particles. In the case where multiple star cluster particles are within a cell, the injected thermal energy is multiplied by the number of star cluster particles. 

Each star cluster particle is expected to produce $N_{\rm SN} = \chi M_{\rm cl} / m_{\rm big}$ SNe, where $\chi$ is the mass fraction of massive stars in the stellar population, $m_{\rm big}$ is the average mass of a massive star in the stellar population, and $M_{\rm cl}$ is the star cluster particle mass. \cite{andalman_origin_2025} use $m_{\rm big} = 10 \; \rm M_\odot$ and $\chi = 0.2$ loosely following a \cite{chabrier_galactic_2003} initial mass function. The number of SNe events per timestep is drawn from a Poisson distribution to match the desired SNe rate. We compute the energy and momentum associated with SNe feedback using power law fitting functions from \citet{martizzi_supernova_2015}, calibrated to their detailed simulations of individual SNe explosions. If the cooling radius is resolved $R_{\rm cool} > 4\Delta x$ then thermal energy $E_{\rm SN} \simeq 10^{51} \; \rm ergs$ is injected into the grid cell. If the cooling radius is not resolved $R_{\rm cool} < \Delta x$ then instead momentum is injected to mimic the "snow plough" phase of the blast wave. In the case where the cooling radius is marginally resolved $\Delta x < R_{\rm cool} < 4 \Delta x$, we interpolate to inject both thermal energy and momentum.

\begin{table*}
    \begin{center}
    \begin{tabular}{l|cccccc}
         Run Name & AGN Feedback & Seed Mass [$\rm M_\odot$] & $\lambda_{\rm edd.}$ & Stellar Feedback \\ 
         \hline
         \texttt{Fiducial\_Andalman+25} & N/A &  N/A & N/A & Fiducial \\
         \color{sim_blue} \texttt{super\_edd\_AGN} & Yes &  $10^4$ & 3 & Fiducial \\
         \color{sim_blue} \texttt{super\_edd\_no\_AGN} & No &  $10^4$ & 3 & Fiducial \\
         \color{sim_green} \texttt{no\_edd\_AGN} & Yes &  $10^4$ & $\infty$ & Fiducial \\
         \color{sim_green} \texttt{no\_edd\_no\_AGN} & No &  $10^4$ & $\infty$ & Fiducial \\
         \color{sim_red} \texttt{edd\_AGN} & Yes &  $10^4$ & 1 & Fiducial \\
         \color{sim_red} \texttt{edd\_no\_AGN} & No &  $10^4$ & 1 & Fiducial \\
         \color{sim_yellow} \texttt{high\_seed\_edd\_AGN}  & Yes &  $10^5$ & 1 & Fiducial \\
         \color{sim_yellow} \texttt{high\_seed\_edd\_no\_AGN} & No &  $10^5$ & 1 & Fiducial \\
         \color{sim_purple} \texttt{low\_stellar\_edd\_AGN} & Yes & $10^4$ & 1 & Low \\
         \color{sim_purple} \texttt{low\_stellar\_edd\_no\_AGN} & No & $10^4$ & 1 & Low \\
    \end{tabular}
    \caption{We present all the numerical experiments discussed in this work. There are a total of 11 simulations where we toggled AGN feedback, varied SMBH seed mass, varied the limit of accretion onto the SMBH, and varied the feedback efficiency. Throughout this work we use the same colour scheme as presented in this table to reference each simulation graphically. In all figures referencing individual simulations, unless otherwise stated, solid coloured curves following the run name colours represent simulations with AGN feedback and dashed coloured curves following the run name colours represent simulations without AGN feedback.}
    \label{tab:sims}
    \end{center}
\end{table*}

\subsection{SMBH Formation \& Growth}
\label{subsec: SMBH formation}

There are 3 proposed formation channels (light seeds, intermediate seeds, and heavy seeds) for a SMBH within a host galaxy (see Section \ref{sec:intro}). While the formation channel of SMBHs is an open question, it is not the focus of this work. We use heavy seed masses $10^4 \; \rm M_\odot$ and $10^5 \; \rm M_\odot$ motivated by observational constraints on seed mass from the local Universe \citep{greene_mass_2007, goulding_towards_2009}. Our default seed mass of $10^4 \; \rm M_\odot$ resides at the boundary between intermediate and heavy seed formation channels. Both of these channels require the formation site to be dense in either gas or stars. For this reason, we use the native on-the-fly clump finder \textsc{phew} \citep{bleuler_phew_2015} within \textsc{ramses} to locate dense ``clumps'' within the simulation. We require that the formation site for a sink to form must have a halo mass of $10^8 \; \rm M_\odot$ with gas and stellar masses of $10^5 \; \rm M_\odot$. In most of our simulations, we choose our initial sink mass representing the SMBH to be $M_{\rm seed} = 10^4 \; \rm M_\odot$ (a tenth of the surrounding mass in stars and gas) which coincidentally similar to the dark matter mass resolution $\simeq m_{\rm dm, min}$. 

Once the SMBH forms, it accretes the surrounding gas within a sphere of fixed radius $r_{\rm sink} = 4 \Delta x_{\rm min} \simeq 40 \; \rm pc$. We use a subgrid model for accretion onto the SMBH, since we cannot resolve the intricate accretion physics near the horizon scale \citep[recent breakthroughs have enabled multi-scale studies that resolve the horizon scale, e.g.][]{hopkins_2024_Forged_in_Fire, guo_2025_cyclic_zoom}. We use the subgrid model described in \cite{biernacki_dynamics_2017}, which employs a modified version of the Bondi-Hoyle-Lyttleton formula \citep{hoyle_effect_1939, bondi_mechanism_1944, bondi_spherically_1952} most applicable to accretion in turbulent flows \citep{krumholz_general_2005}. The formula for the Bondi accretion rate is written as
\begin{equation}
    \dot{M}_{\rm bondi} = 4 \pi \rho_\infty r^2_{\rm bondi} v_{\rm bondi} \; ,
\label{eq:mdot_bondi}
\end{equation}
where we define the Bondi velocity $v_{\rm bondi}$ and the Bondi radius $r_{\rm bondi}$ to be 
\begin{align}
    v_{\rm bondi} = \sqrt{c_{\rm s, eff}^2 + v_{\rm rel}^2}, \quad r_{\rm bondi} = \frac{G M_{\rm sink}}{v_{\rm bondi}^2} \; .
\end{align}
In this work we ignore the relative velocity $v_{\rm rel}$ between the gas and the sink so that the Bondi velocity is only dependent on the effective sound speed $c_{\rm s, eff}$. The effective sound speed is equal to the sound speed within the sink radius $c_{\rm s}$ divided by a boost factor $\beta_{\rm boost}(\rho) = \max\left[ (\rho/\rho_\star)^{2/3}, 1.0\right]$ which depends on a threshold critical gas density $\rho_\star$ which we set to be 1 H/cc following \citet{biernacki_dynamics_2017}. We boost the accretion rate to account for unresolved density and temperature fluctuations smaller than $\Delta x_{\rm min} = 10 \; \rm pc$. This treatment of boosting the accretion rate is equivalent to the ubiquitous boosting treatment of \cite{booth_cosmological_2009}. Lastly, $\rho_\infty$ is defined as
\begin{equation}
    \rho_\infty = \frac{\langle \rho \rangle}{\alpha_{\rm B} (x_{\rm sink})} \;,
\end{equation}
where $\langle \rho \rangle$ is the average gas density within the sink sphere, $\alpha_{\rm B}$ is the dimensionless density profile of a Bondi accretion flow, and $x_{\rm sink} = r_{\rm sink} / r_{\rm bondi}$ is the dimensionless sink radius. The dimensionless density profile $\alpha_{\rm B}(x)$ approaches $\alpha_{\rm B} \propto x^{-3/2}$ when the flow within the sink radius is strongly supersonic ($x_{\rm sink} \ll 1$ and it approaches $\alpha_{\rm B} \simeq 1$ when the flow within the sink radius is strongly subsonic ($x_{\rm sink} \gg 1$)  \citep{krumholz_general_2005}. The boost factor $\beta_{\rm boost}$ alters where the effective transition from subsonic to supersonic lies in this formalism \citep{biernacki_dynamics_2017}. 

Since we do not self-consistently model radiation, we impose a limit on the Bondi accretion formula where radiation pressure has a significant effect on the accretion flow. This maximum limit on the accretion rate is commonly known as the Eddington rate 
\begin{equation}
    \dot{M}_{\rm edd.} = \frac{4 \pi G M_{\rm sink} m_{\rm p}}{\epsilon_{\rm r} \sigma_{\rm T} c} \; ,
\label{eq:mdot_edd_full}
\end{equation}
where $m_{\rm p}$ is the proton mass, $\sigma_{\rm T}$ is the Thomson cross section, and $\epsilon_{\rm r} = 10\%$ is the radiative efficiency proposed by \cite{shakura_black_1973}.\footnote{We note that the radiative efficiency could be significantly lower than $10\%$ \citep{sadowski_2014_grmhd_low_rad, ryan_2017_grmhd_low_rad, davies_2019_low_rad} which could raise the Eddington rate. Lowering this efficiency would have the same effect as raising the super-Eddington accretion factor $\lambda_{\rm edd.}$ in our simulations.} When accreting at the Eddington rate, the SMBH growth is exponential 
\begin{equation}
    \dot{M}_{\rm edd.} = \frac{M_{\rm sink}}{\tau_{\rm sal}}
\label{eq:mdot_edd_simple}
\end{equation}
with a characteristic Salpeter timescale $\tau_{\rm sal} \simeq 45 \; \rm Myr$. We define the maximal accretion factor $\lambda_{\rm edd.} \equiv \dot{M}^{\rm max}_{\rm acc} / \dot{M}_{\rm edd.}$ to increase the Eddington rate by a factor of  $\lambda_{\rm edd.}$ allowing super-Eddington accretion in our simulations. The standard Eddington limit occurs when we set $\lambda_{\rm edd.} = 1$. 

Ultimately, the accretion rate of the SMBH is given by 
\begin{equation}
    \dot{M}_{\rm acc} = \min \left[\dot{M}_{\rm bondi}, \lambda_{\rm edd.} \dot{M}_{\rm edd.} \right] \; .
\label{eq:mdot_acc}
\end{equation}
In practice we divide the Salpeter timescale by $\lambda_{\rm edd.}$ to alter the maximal accretion rate. As the SMBH accretes gas from within the sink sphere, the sink sphere loses gas mass $\Delta M_{\rm gas} = -\dot{M}_{\rm acc} \Delta t$. This is done by removing density from each cell in the sink sphere with mass-weighting for each cell to prevent completely stripping cells of low density. 

\subsection{SMBH Feedback}
\label{subsec: SMBH feedback}

As the SMBH grows, it releases energy into the surrounding environment. In the high redshift Universe, observations suggest AGN feedback is dominated by the ``quasar mode'': a phase of feedback characterized by a strong isotropic AGN-driven wind that quickly thermalizes the surrounding nuclear environment through forward and reverse shocks. Therefore, in our simulations we inject purely thermal energy into the sink sphere isotropically. The luminosity of our SMBH is given by 
\begin{equation}
    L_{\rm AGN} = \epsilon_{\rm r} \epsilon_{\rm c} \dot{M}_{\rm acc} c^2 \; ,
\label{eq:L_agn}
\end{equation}
where $\epsilon_{\rm c}$ is the thermal energy coupling efficiency parameter. The energy from the SMBH is deposited at every time step $\Delta E_{\rm AGN} = L_{\rm AGN} \Delta t$ into the cells of the sink sphere with mass-weighting.

In cosmological simulations, $\epsilon_{\rm c}$ is typically chosen between $5\%$ and $15\%$ depending on resolution to match observables\citep{springel_simulations_2005, booth_cosmological_2009, wurster_comparative_2013,  gabor_simulations_2013}. In our simulations, we consider strong AGN feedback $\epsilon_{\rm c} = 15\%$ and no AGN feedback $\epsilon_{\rm c} = 0\%$, which lets us isolate its effect.

This implementation of AGN feedback is simplistic relative to the state-of-the-art subgrid modelling for super-Eddington SMBHs in cosmological zoom-in simulations \citep{massonneau_how_2023, lupi_sustained_2024, husko_effects_2025}. Flows in the super-Eddington regime can depart significantly in geometry, radiative efficiency, and outflow properties from the assumed thin-disc flow. By assuming the same coupling efficiency across accretion regimes, we may under-predict the ability of the AGN to self-regulate during periods of maximal accretion, biasing growth towards the super-Eddington regime in our simulations.

Models that self-consistently track SMBH spin \citep{lupi_sustained_2024, husko_effects_2025} find that SMBHs can achieve low spins relative to expectations, resulting in lower feedback efficiencies and sustained super-Eddington accretion. Our prescription also ignores the role of jets (kinetic feedback), which can play a significant role in altering the amount of time the SMBH spends in the self-regulated regime \cite{massonneau_how_2023}. Future work is needed to explore these effects and their role in SMBH growth in the most massive galaxies at Cosmic Dawn.

\subsection{Simulation Suite}
\label{subsec: sim suite}

We run 11 simulations to investigate different SMBH growth scenarios within an MDG at Cosmic Dawn. Our suite of simulations are summarized in Table \ref{tab:sims}. We vary 3 aspects of SMBH growth within the simulation: strength of AGN feedback $\epsilon_{\rm c} = (0\%, 15\%)$, the maximal growth rate $\lambda_{\rm edd.} = (1, 3, \infty)$, and the seed mass of SMBH $M_{\rm seed} = (10^4, 10^5) \; \rm M_\odot$. Across all of our simulations we hold the same seeding criteria that $M_{\rm gas}, M_{\star} \geq 10^5 \; \rm M_\odot$ and $M_{\rm halo} \geq 10^8 \; \rm M_\odot$ which corresponds to a seeding time of $z \sim 15$ within the host MDG. 

As a point of reference, we reproduce the fiducial MDG simulation of \citet{andalman_origin_2025} with no sink particles. Additionally, we include simulations with lower stellar feedback than the fiducial model to explore the impact of stellar feedback on SMBH growth within the MDG. We achieve this by removing early-type feedback (equivalent to \texttt{SNeOnly} or \texttt{noPhot} simulations of \citet{andalman_origin_2025}), leaving only feedback from core-collapse SNe. In the absence of photoionizing feedback the cooling radius $R_{\rm cool}$ remains unresolved which means that SNe can only inject momentum into the ISM. We refer to this feedback model as the low stellar feedback model in Table \ref{tab:sims}. The low stellar feedback model creates a thinner and colder disk than the fiducial model, allowing for more efficient SMBH growth.

\begin{figure*}
    \centering
    \includegraphics[width=\linewidth]{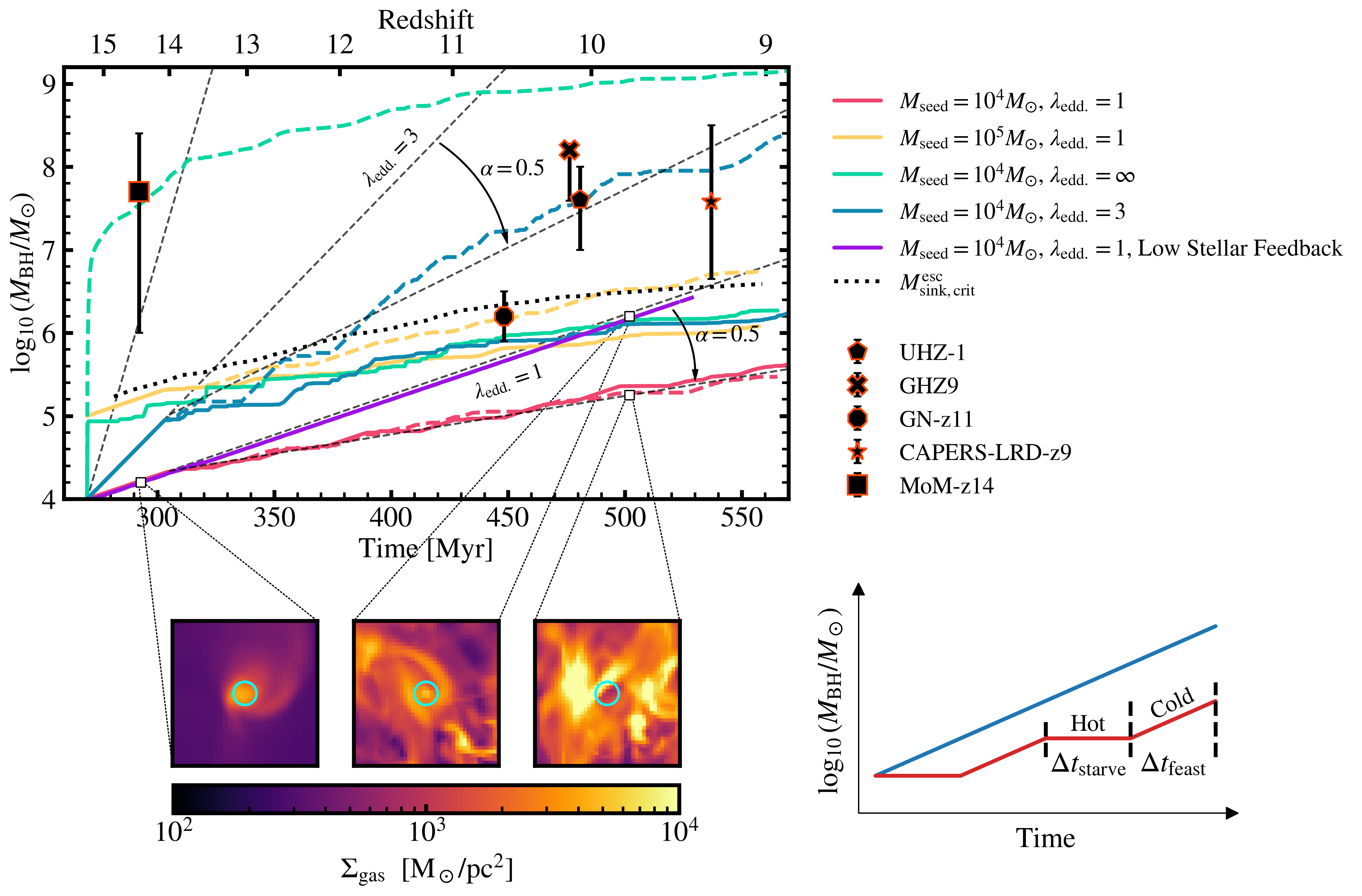}
    \caption{Mass evolution of the SMBH as a function of time for all simulations in our suite (see Table \ref{tab:sims}). Solid coloured curves denote simulations with AGN feedback and dashed coloured curves denote simulations without AGN feedback. Thin-black dashed lines correspond to exponential (super-)Eddington growth with a slope determined by $\lambda_{\rm edd.}$. The black dotted curve denotes the critical self-regulated mass $M^{\rm esc}_{\rm sink,crit}$ required for AGN feedback to unbind gas in the sink sphere from the halo. For context, we include observations of Cosmic Dawn AGN with \textit{JWST} \citep{maiolino_small_2024, natarajan_first_2024, napolitano_dual_2025, taylor_capers-lrd-z9_2025, naidu_black_2025}. The zoom-in panels show snapshots of the projected density field within a 500 pc region around the sink sphere at early and late times for different simulations. The cyan rings denote the sink sphere with radius $r_{\rm sink} = 40 \; \rm pc$. The arced arrows denote the suppression in the growth slope caused by the turbulent-multiphase ISM environment. We provide a heuristic diagram to show how the environment impacts SMBH growth in the bottom right. Simulations with no AGN feedback in the fiducial ISM model show the same amount of starvation ($\alpha \approx 0.5$) while simulations with the low stellar feedback model indicate no starvation ($\alpha = 1$).}
    \label{fig:SMBH_mass_vs_time}
\end{figure*}

We find that the simulations with $M_{\rm seed} = 10^4 \; \rm M_\odot$, $\lambda_{\rm edd.} = 1$ are the only simulations in our suite where AGN feedback strength does not play a significant role in SMBH evolution (see Fig. \ref{fig:SMBH_mass_vs_time}). This finding motivates the inclusion of simulations varying the maximal growth rate and the seed mass as we expect these parameters to have the most dominant impact on the resulting SMBH mass by $z \sim 9$. Beyond the Eddington limit, we explore mildly super-Eddington accretion $\lambda_{\rm edd.} = 3$ and extreme super-Eddington accretion $\lambda_{\rm edd.} = \infty$ where we remove the Eddington limit entirely. In our simulations, extreme super-Eddington accretion achieves a maximum accretion rate comparable to $\lambda_{\rm edd.} \sim 10^3 - 10^4$. Both super-Eddington simulations show a significant difference in SMBH evolution when AGN feedback strength is varied. The super-Eddington simulations with AGN feedback deviate from their no AGN feedback counterpart when $M_{\rm sink} \sim 10^5 \; \rm M_\odot$, this motivates our choice of seed mass for the {\texttt{high\_seed\_edd\_AGN}} and {\texttt{high\_seed\_edd\_no\_AGN}} simulations. In Section \ref{sec:results} we explain the diversity in SMBH evolution between our simulations and explore the impact of the SMBH on the host galaxy across all of the simulations.

\section{Results}
\label{sec:results}
In Section \ref{subsec:regimes} we first present an analytical framework to understand the growth of the SMBH across our simulations. By modelling the competition between heating and cooling in the sink sphere we gain insights into transitions between different accretion regimes. We use this modelling framework to explain how the SMBH grows within the host galaxy across all our simulations in Section \ref{subsec:SMBH Growth}. In Section \ref{subsec:impact} we describe the impact of the SMBH on the host galaxy across all of our simulations. In Section \ref{subsec:timescales} we identify relevant resolution-independent growth timescales which offer physical insights about SMBH growth in massive galaxies at Cosmic Dawn. 

\subsection{SMBH Accretion Regimes}
\label{subsec:regimes}

Following \citet{biernacki_dynamics_2017}, SMBH growth in our simulations can be understood by considering the competition between heating and cooling within the sink sphere. Gas near the SMBH is heated by AGN feedback and cooled by radiative losses. The mean specific internal energy within the sink sphere is governed by the following equation
\begin{equation}
    \rho \frac{d\epsilon}{dt} = \underbrace{\frac{L_{\rm AGN}}{V_{\rm sink}}}_{\textup{AGN Heating}} - \;\;\; \underbrace{n_{\rm H}^2 \Lambda(T, Z)}_{\textup{Cooling}} \; ,
    \label{eq:epsdot}
\end{equation}
where $V_{\rm sink}$ is the sink sphere volume, $n_{\rm H}$ is the gas density, and $\Lambda(T, Z)$ is the cooling function, which depends on temperature and metallicity \citep{katz_cosmological_1996, sutherland_cooling_1993}. The specific internal energy is related to the temperature and the sound speed by 
\begin{equation}
    \epsilon \simeq \frac{k_{\rm B}T}{\mu m_{\rm H}} \simeq c_{\rm s}^2(t) \; ,
    \label{eq:eps_to_cs}
\end{equation}
where we have assumed an ideal gas.

Like \citet{biernacki_dynamics_2017}, we identify two accretion regimes: cold and hot. In the cold accretion regime, the cooling term dominates the heating term in equation~\ref{eq:epsdot}. As a result, the sound speed is kept close to the floor temperature. If the SMBH is at rest in the centre of the galaxy, then the Bondi accretion rate is
\begin{equation}
    \dot{M}_{\rm bondi} \simeq 4\pi \rho \frac{(G M_{\rm sink})^2}{c_{\rm s}^3(t)} \; .
\end{equation}
Therefore, a low sound speed implies $\dot{M}_{\rm bondi} \gg \lambda_{\rm edd.} \dot{M}_{\rm edd.}$, so the accretion rate is limited by $\lambda_{\rm edd.} \dot{M}_{\rm edd.} \propto M_{\rm sink}$. If the SMBH accretes at this maximum rate, then it will grow exponentially with a characteristic growth timescale of $\tau_{\rm sal} \simeq 45 \; \mathrm{Myr} / \lambda_{\rm edd.}$. In Fig. \ref{fig:SMBH_mass_vs_time} we show these exponential growth tracks as black dashed curves.

\subsubsection{Transition from Cold to Hot Accretion Regime} 

As the SMBH grows, the Eddington accretion rate increases and heating from AGN feedback becomes increasingly important relative to radiative cooling. The SMBH transitions into the hot accretion regime when the heating and cooling terms are equal. 
This transition occurs at a critical SMBH mass
\begin{align}
    M_{\rm sink, crit}^{\rm cool} = \frac{\tau_{\rm sal}}{\lambda_{\rm edd.}\epsilon_{\rm c}\epsilon_{\rm r}c^2} n_{\rm H}^2 \Lambda(T, Z)  V_{\rm sink}  \; .
\label{eq:Mcool_explicit}
\end{align}
We are interested in the critical mass corresponding to the maximum (over temperature) of the cooling function $\Lambda(T,Z)$. which we parametrize as 
\begin{equation}
\Lambda(Z) \simeq \mathrm{max}\left[\Lambda(T,Z)\right] \simeq 
    \begin{cases}
    5\times10^{-22} \left(\frac{Z}{Z_\odot} \right),& \text{if } Z\geq 0.2 \; Z_\odot\\
    10^{-22},              & \text{otherwise}
    \end{cases}
     \; .
\end{equation}
This parametrization is most applicable to gas with significant metal enrichment. If $M_{\rm sink} > M_{\rm sink, crit}^{\rm cool}$ when $\Lambda \simeq  \mathrm{max}\left[\Lambda(T,Z)\right]$ then heating dominates over cooling regardless of the temperature of the gas. In other words, if AGN heating causes the temperature to rise within the sink sphere after cooling has reached a maximum, then there is no way for cooling to stop runaway heating. 

\begin{figure}
    \centering
    \includegraphics[width=\linewidth]{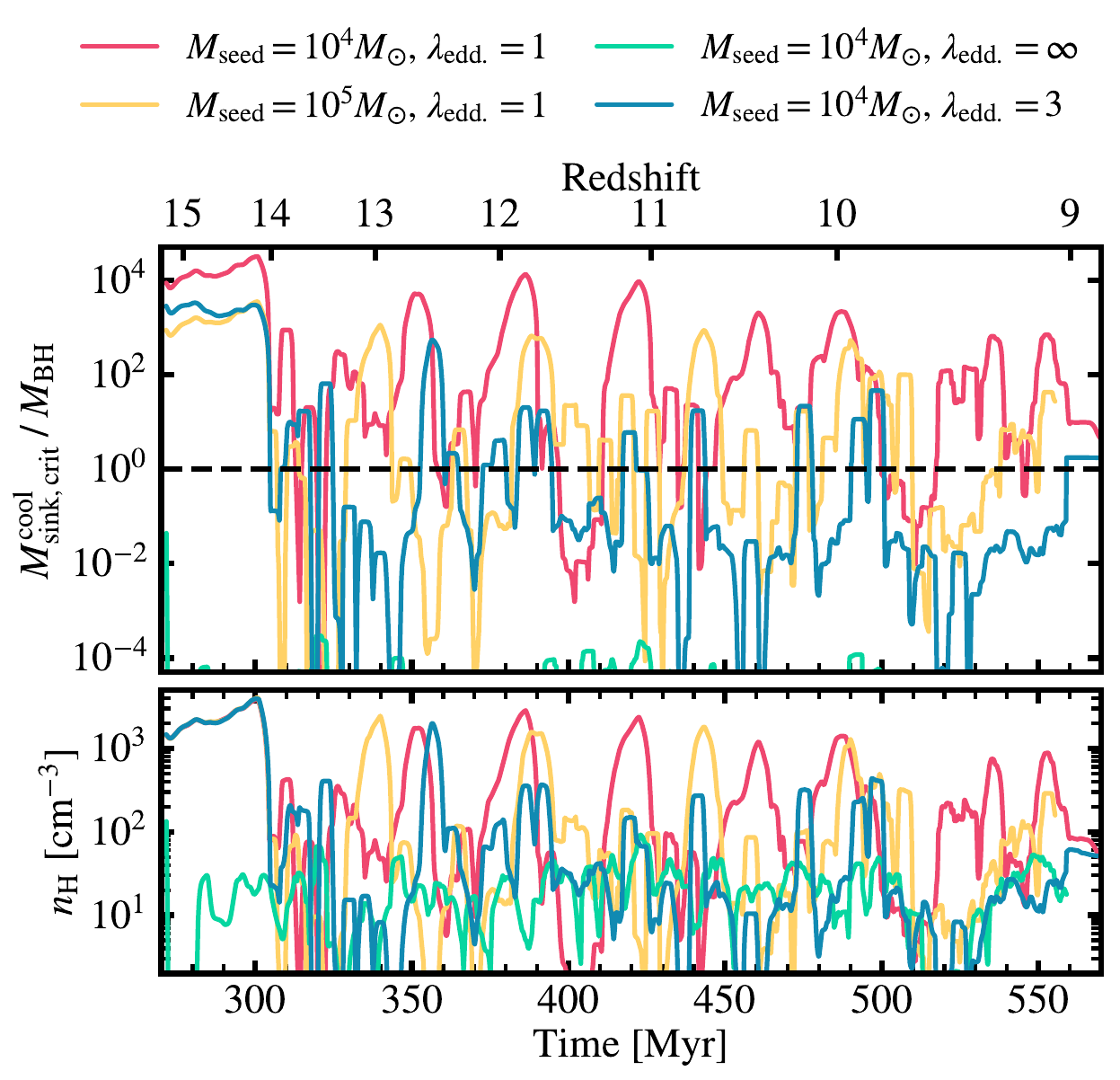}
    \caption{In the top panel we show the critical mass threshold for heating from AGN feedback to dominate over cooling $M^{\rm cool}_{\rm sink,crit}$ relative to the sink mass as a function of time for all simulations with AGN feedback enabled. In the bottom panel we show the gas density within the sink sphere for each simulation. For clarity, all curves shown are running averages with a window size of 5 Myrs.}
    \label{fig:mass_thresholds}
\end{figure}

\begin{figure*}
    \centering
    \includegraphics[width=\linewidth]{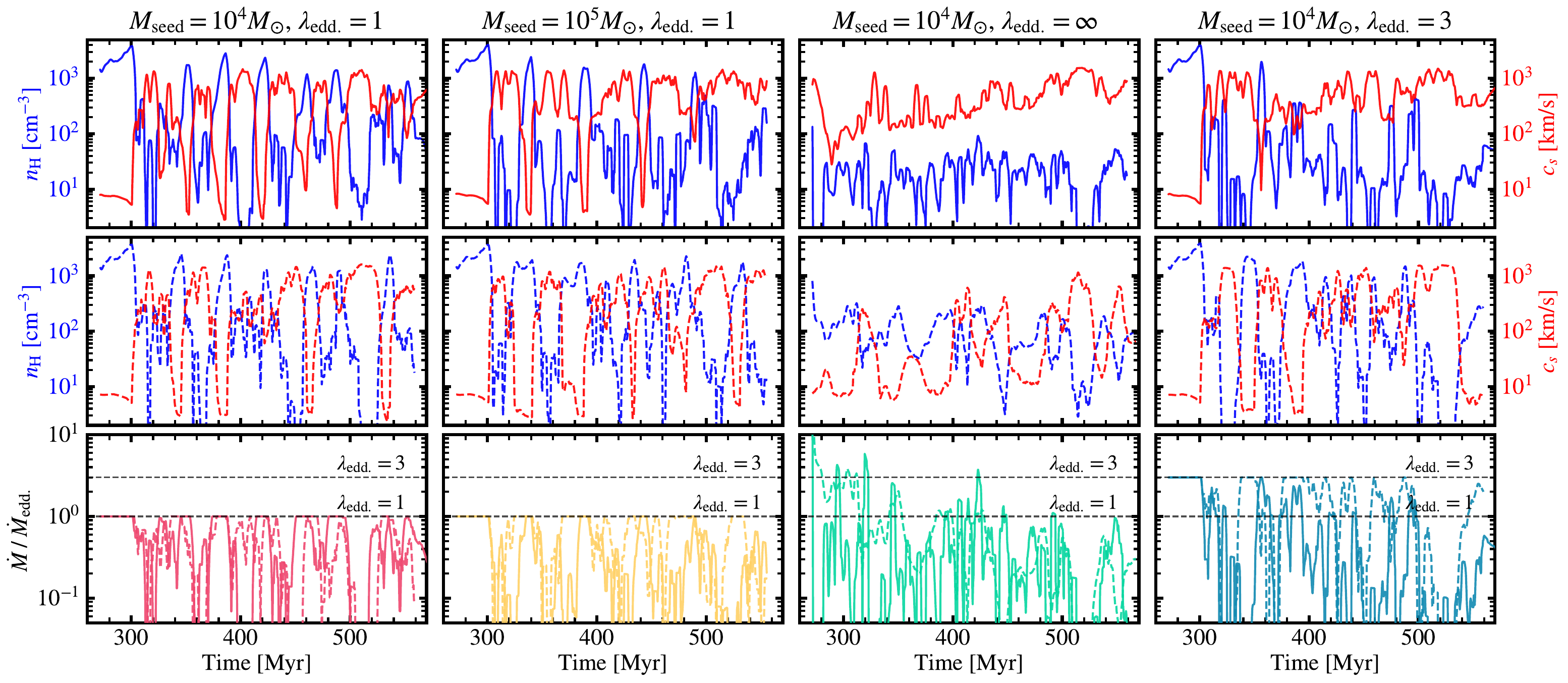}
    \caption{Sink sphere properties as a function of time. Each column represents a different set of simulations from the suite outlined in Table \ref{tab:sims}. In all panels, solid lines denote simulations with AGN feedback and dashed lines denote simulations without AGN feedback. The top and middle rows show the density (blue) and sound speed (red) within the sink sphere. The bottom row shows the accretion rate relative to the Eddington rate. For clarity all curves shown are running averages with a smoothing window of 5 Myrs.}
    \label{fig:sink_sphere_env}
\end{figure*}

Plugging in our parametrized version of the cooling function we have
\begin{equation}
    M_{\rm sink, crit}^{\rm cool}\simeq \frac{2 \times 10^4 \; \mathrm{M_\odot}}{\lambda_{\rm edd.}} \left( \frac{n_{\rm H}}{10 \; \rm cm^{-3}} \right)^2 \left( \frac{Z}{Z_\odot} \right) \left(\frac{r_{\rm sink}}{40 \; \rm pc} \right)^3 \; .
    \label{eq:Mcool}
\end{equation}
We show the time evolution of this threshold in the top panel of Fig. \ref{fig:mass_thresholds}. To account for metal enrichment over time we use the average mass-weighted metallicity within the sink sphere as a function of time in our calculation of the critical mass threshold.\footnote{We do not store metallicity in the sink sphere at every time step, but instead at every output dump. We linearly interpolate the metallicity between outputs to include it in our calculation of the critical cooling mass threshold.} An important departure from the work of \cite{biernacki_dynamics_2017} comes from the difference in the underlying nature of the ISM. In the MDG environment, the turbulent-multiphase ISM consists of sloshing patches of cold-dense ($n_{\rm H} \sim 10^3 \; \rm cm^{-3} , c_{\rm s} \sim 10 \; \rm km/s$) and hot-diffuse ($n_{\rm H} \sim 10 \; \rm cm^{-3} , c_{\rm s} \sim 10^3 \; \rm km/s$) gas with a density and sound speed contrast between the two phases of $\gtrsim 2-3$ orders of magnitude (see Fig. \ref{fig:sink_sphere_env}). In \cite{biernacki_dynamics_2017}, the SMBH undergoes sustained accretion of dense gas from a relatively stable ISM in a late-type galaxy. 

In the late type galaxy environment of \cite{biernacki_dynamics_2017} the SMBH only has to exceed the critical mass once to transition into the heating dominated regime and achieve self-regulation from runaway heating. However, in the MDG environment, the SMBH often exceeds the critical mass in hot-diffuse pockets of the ISM, but not in cold-dense patches where the critical mass is larger by a factor up to $10^4$ (Fig. ~\ref{fig:mass_thresholds}).

When the SMBH sits in a cold-dense patch of the ISM, it enters deep into the cold accretion regime and accretes maximally at the (super-)Eddington-limited rate, which we call the feast mode. When the SMBH enters a hot-diffuse pocket, it enters into the hot accretion regime and accretes at the Bondi-limited rate ($\dot{M}_{\rm bondi} \ll \lambda_{\rm edd.} \dot{M}_{\rm edd.}$), which we call the starvation mode. The naming reflects the fact that the accretion rate is so small in the starvation mode that the SMBH growth effectively stagnates.\footnote{Throughout this work, any reference to the feast and starvation modes of accretion implies environment-induced entry into the cold and hot accretion regime.} We provide a schematic of this picture in the bottom right panel of Fig. \ref{fig:SMBH_mass_vs_time}. The underlying properties of the ISM have a direct impact on the growth of the SMBH by determining the duration and stochasticity of feast and starvation modes (Section \ref{subsec:SMBH Growth}). 

If the mass of the SMBH is greater than the critical cooling mass threshold for the majority of the time spent in hot-diffuse pockets of the ISM ($\gtrsim 50\%$), then AGN feedback will influence the surrounding environment of the SMBH (Section \ref{subsec:SMBH Growth}). If the SMBH is not massive enough to sustain being above the critical cooling mass threshold, then it may not become capable of generating enough thermal pressure to influence the surrounding environment and achieve self-regulation at Cosmic Dawn. 

\subsubsection{Hot Accretion Regime to Self-Regulation} 

If we assume that the SMBH has been in the hot accretion regime long enough to mediate the nuclear region of the galaxy, then we can ignore the cooling term in equation \ref{eq:epsdot} and take $\dot{M}_{\rm acc} = \dot{M}_{\rm bondi}$. We then solve the simplified differential equation for the time evolution of the sound speed assuming Bondi-limited accretion of an ideal gas onto the SMBH, as done in \cite{biernacki_dynamics_2017}, which yields 
\begin{equation}
    c_{\rm s}(t) = \left[ \frac{15}{2} \epsilon_{\rm c} \epsilon_{\rm r} c^2 \left( \frac{G M_{\rm sink}}{r_{\rm sink}} \right)^2 \frac{t}{r_{\rm sink}} \right]^{1/5} \; .
    \label{eq: sound_speed_vs_t}
\end{equation}
This model implies that the temperature in the sink region grows indefinitely. However, eventually the gas near the SMBH expands and cools adiabatically. The relevant timescale for cooling via adiabatic expansion is the sound-crossing time of the sink sphere $\tau_{\rm cross}(t) = r_{\rm sink} / c_{\rm s}(t)$. By plugging this timescale into equation \ref{eq: sound_speed_vs_t}, we find the maximum sound speed in the hot accretion phase is
\begin{equation}
    c_{\rm s, max} = c_{\rm s}(\tau_{\rm cross}) = \left[ \frac{15}{2} \epsilon_{\rm c} \epsilon_{\rm r} c^2 \left( \frac{G M_{\rm sink}}{r_{\rm sink}} \right)^2 \right]^{1/6}
\end{equation}

If the SMBH is massive enough such that $c_{\rm s, max}$ exceeds the escape velocity of the halo $v_{\rm esc} = \sqrt{2GM_{\rm halo} / r_{\rm halo}}$, then it is possible for the SMBH to unbind gas from the halo via runaway thermal heating from AGN feedback. By equating $v_{\rm esc}$ to $c_{\rm s,max}$, we find the critical SMBH threshold mass is
\begin{equation}
    M_{\rm sink, crit}^{\rm esc} = \frac{r_{\rm sink}}{G} \sqrt{\frac{2}{15} \frac{1}{\epsilon_{\rm c} \epsilon_{\rm r} c^2} \left( \frac{2 G M_{\rm halo}}{r_{\rm halo}} \right)^3} \; ,
    \label{eq:Mesc}
\end{equation}
where the halo radius is given by
\begin{equation}
    r_{\rm halo} = \left( \frac{3}{4\pi} \frac{M_{\rm halo}}{\Delta \overline{\rho}_{\rm m}} \right)^{1/3} \; .
\label{eq:r_halo}
\end{equation}
where $\overline{\rho}_{\rm m}$ is the mean comoving density of the universe and $\Delta = 200$ is the critical overdensity for spherical collapse. If $M_{\rm sink} > M_{\rm sink, crit}^{\rm esc}$ then we expect the thermal energy deposited into the sink sphere by AGN feedback to be strong enough to completely remove any gas that could be accreted, effectively regulating the growth of the SMBH. Therefore, in the absence of stellar feedback, $M_{\rm sink, crit}^{\rm esc}$ represents the self-regulated mass of the SMBH. Numerically this self-regulated mass can be written as 
\begin{equation}
    M_{\rm sink, crit}^{\rm esc} \simeq 4 \times 10^6 \; \mathrm{M_\odot} \left( \frac{r_{\rm sink}}{40 \; \rm pc} \right) \left( \frac{M_{\rm halo}}{10^{11} \; \rm M_\odot} \right) \; ,
\end{equation}
which corresponds to the self-regulated mass at the end of our simulation when $z = 9$. In our simulations, stellar feedback complicates this picture by seeding the turbulent-multiphase ISM, so $M_{\rm sink, crit}^{\rm esc}$ is an effective upper bound on the mass of the SMBH in the hot regime which can be seen in Fig. \ref{fig:SMBH_mass_vs_time} (black-dotted curve).

Both critical masses are explicit functions of $r_{\rm sink}$, with $M_{\rm sink,crit}^{\rm cool} \propto r_{\rm sink}^3$ and $M_{\rm sink,crit}^{\rm esc} \propto r_{\rm sink}$, meaning they are explicitly resolution dependent quantities. If our effective minimum resolution was $1 \; \rm pc$ instead of $10 \; \rm pc$ then we would expect the cold-to-hot accretion regime transition mass to be $1000\times$ smaller and the self-regulated mass to be $10\times$ smaller. This is obviously not a satisfactory result, as it prevents our model to be fully predictive. One can speculate that multiple missing physical ingredients could provide fixed physical scales over which feedback energy will be deposited (non-equilibrium radiation, relativistic particles and cosmic rays, magnetic fields...) but this is beyond the scope of this paper.

In summary, we have described the possible states the SMBH can be in throughout the simulation. The possible accretion regimes are bifurcated into the cooling-dominated (cold) maximal accretion regime limited by $\lambda_{\rm edd.}\dot{M}_{\rm edd.}$ and the heating-dominated (hot) sub-maximal accretion regime limited by $\dot{M}_{\rm bondi}$. The transition between these two regimes occurs when $M_{\rm sink} \simeq M^{\rm cool}_{\rm sink,crit}$. The ISM in the MDG environment is turbulent and multiphase with cold-dense and hot-diffuse patches which the sink sphere can plunge into, complicating this picture by raising or lowering $M^{\rm cool}_{\rm sink,crit}$ by $\sim \pm 4$ orders of magnitude, triggering feast or starvation modes of accretion. If $M_{\rm sink} > M^{\rm cool}_{\rm sink,crit}$ for sufficiently long, then the thermal pressure from AGN feedback will be strong enough to influence the surrounding environment, effectively stabilizing the ISM immediately surrounding the sink sphere. If this happens then there will be sufficient time for runaway AGN heating that can unbind gas from the halo. The mass of the SMBH then grows in step with the self-regulated mass $M_{\rm sink, crit}^{\rm esc}$. In Section \ref{subsec:SMBH Growth} we examine the SMBH growth across all simulations in our suite with this framework in mind.

\subsection{Growth of the SMBH}
\label{subsec:SMBH Growth}

All simulations in our suite (outlined in Table \ref{tab:sims}) with the fiducial stellar feedback model have 2 distinct epochs of growth: the calm accretion epoch limited by $ \lambda_{\rm edd.} \dot{M}_{\rm edd.}$ and the bursty accretion epoch. In Fig. \ref{fig:SMBH_mass_vs_time}, we show that the transition between these two epochs of growth occurs shortly after $t\sim 300~{\rm Myr}$ for all simulations except those where $\lambda_{\rm edd.} = \infty$. One can see this transition as a departure from the black dashed lines denoting exponential (super-)Eddington limited growth. This transition occurs because of stellar feedback. 

\subsubsection{The Bursty Epoch} 

The first significant burst of star formation in our simulated MDGs occurs at $300~{\rm Myr}$ (Sec.~\ref{subsec:impact}). The ensuing stellar feedback drives ISM turbulence and creates pockets of hot-diffuse gas in an otherwise cold-dense medium. The sink sphere stochastically samples this multi-phase structure, leading to alternating periods of feast and starvation mode accretion. 

Stochastic plunges into the cold-dense medium can be seen in the fluctuations of density and sound speed within the sink sphere in Fig. \ref{fig:sink_sphere_env}. This effect is most clear in simulations without AGN feedback (dashed lines) because there is no way for the SMBH to influence the surrounding environment aside from accreting gas. It is clear from the bottom row of Fig. \ref{fig:sink_sphere_env} that when density abruptly rises and the sound speed abruptly falls as the sink sphere enters a cold-dense patch of the ISM, the accretion rate reaches the (super-)Eddington rate (feast mode). Similarly, when the density abruptly falls and the sound speed abruptly rises as the sink sphere leaves a cold-dense patch of the ISM, the accretion rate drops substantially (starvation mode). 

We model the mass evolution that arises from this bursty accretion history as a piecewise exponential function schematically drawn in the bottom-right panel of Fig. \ref{fig:SMBH_mass_vs_time}. The mass evolution after $t \sim 300 \; \rm Myr$ consists of many bursts of maximal exponential accretion followed by short stagnations of growth. The stochasticity of these feast and starvation modes is set by the stochasticity of the SFR. According to Fig. 12 of \cite{andalman_origin_2025}, the power spectral density (PSD) of the SFR follows a red noise distribution over fluctuation timescales of 1 - 10 Myrs and white noise outside of those timescales. The duration and occurrence rates of feast modes reflect these red-noise distributed timescales. Stochasticity in SFR along with the strength of stellar feedback directly sets the distribution of occurrence rates and durations of feast and starvation modes of accretion by setting the properties of the multiphase ISM. 

A common feature seen in all (super-)Eddington limited simulations that use our fiducial stellar feedback model but do not have AGN feedback is that there is a consistent suppression in the exponential growth timescale by a factor 
\begin{equation}
    \alpha \equiv \frac{\int I(t) \; dt}{\int dt}
\label{eq:alpha}
\end{equation}
where $I(t)$ is an indicator function that can be written as
\begin{equation}
    I(t) = 
    \begin{cases}
    1,& \text{if } \dot{M}_{\rm acc} = \lambda_{\rm edd.} \dot{M}_{\rm edd.}\\
    0,              & \text{otherwise}
    \end{cases} \; .
\end{equation}
We find the suppression factor (akin to the duty cycle of the AGN) to be $\alpha \approx 0.5$ for our fiducial ISM model. This implies that the SMBH spends nearly the same amount of time in the starvation mode as it does in the feast mode. The accretion rate in the bursty accretion epoch can be simply described then as $\dot{M}_{\rm acc} \approx \alpha \lambda_{\rm edd.}\dot{M}_{\rm edd.}$ in the absence of AGN feedback.

We show that the suppression factor depends sensitively on the conditions of the ISM by showing the mass evolution of the SMBH in the same MDG but instead with the low stellar feedback model. In the low stellar feedback model early-type stellar feedback is turned off so that SNe are only injecting momentum into the gas. In the low stellar feedback model the ISM does not contain hot-diffuse patches so the density in the nuclear environment is sustained at $n_{\rm H} \sim 10^3 \; \rm cm^{-3}$. This environment allows the SMBH to steadily grow at a maximal accretion rate until it surpasses $M^{\rm cool}_{\rm sink,crit} \sim 10^8 \; \rm M_\odot$ (see equation \ref{eq:Mcool}) which means the SMBH is in the feast mode the entire duration of the simulation ($\alpha = 1$). In Fig.~\ref{fig:SMBH_mass_vs_time}, we compare the projected density field in the vicinity of the sink sphere to illuminate this description. At early times ($t \lesssim 300 \; \rm Myr$) there is no significant difference between stellar feedback models and the cold-dense gas is smoothly varying within the nuclear region. At late times ($t \sim 500 \; \rm Myr$) there are significant differences in the nuclear region of the ISM between the low and fiducial stellar feedback models. The gas density in the fiducial stellar feedback model is patchy with high density contrast between ISM phases while the gas density in the low stellar feedback model is more reminiscent of early times. 

\subsubsection{Self-Regulation} 

We see in the top panel of Fig. \ref{fig:SMBH_mass_vs_time} that once the {\texttt{edd\_AGN}} and {\texttt{edd\_no\_AGN}} simulations enter into the bursty accretion epoch there is no significant difference in SMBH growth between the simulation with no AGN feedback (red dashed curve) and the simulation with AGN feedback (red solid curve). This implies that the SMBH is not self-regulated by AGN feedback but instead entirely regulated by stellar feedback during this epoch. This is in contrast to all other simulations with the fiducial stellar feedback model where we do see a divergence in mass evolution between simulations with AGN feedback and simulations without AGN feedback in the bursty accretion epoch. 

When we examine Fig. \ref{fig:SMBH_mass_vs_time} for differences between the {\texttt{edd\_AGN}} simulation and the rest of the simulations that achieve self-regulation, it is immediately clear that by the time of entry into the bursty accretion epoch, the SMBH in the {\texttt{edd\_AGN}} simulation is an order of magnitude less massive than in the other simulations. One can intuit that the difference in SMBH mass at this point in time is crucial to explain this behaviour. Using the formalism described in Section \ref{subsec:regimes}, we can understand why by comparing the sink mass to the critical cooling mass $M^{\rm cool}_{\rm sink, crit}$ as done in Fig. \ref{fig:mass_thresholds}. 

\begin{figure}
    \centering
    \includegraphics[width=\linewidth]{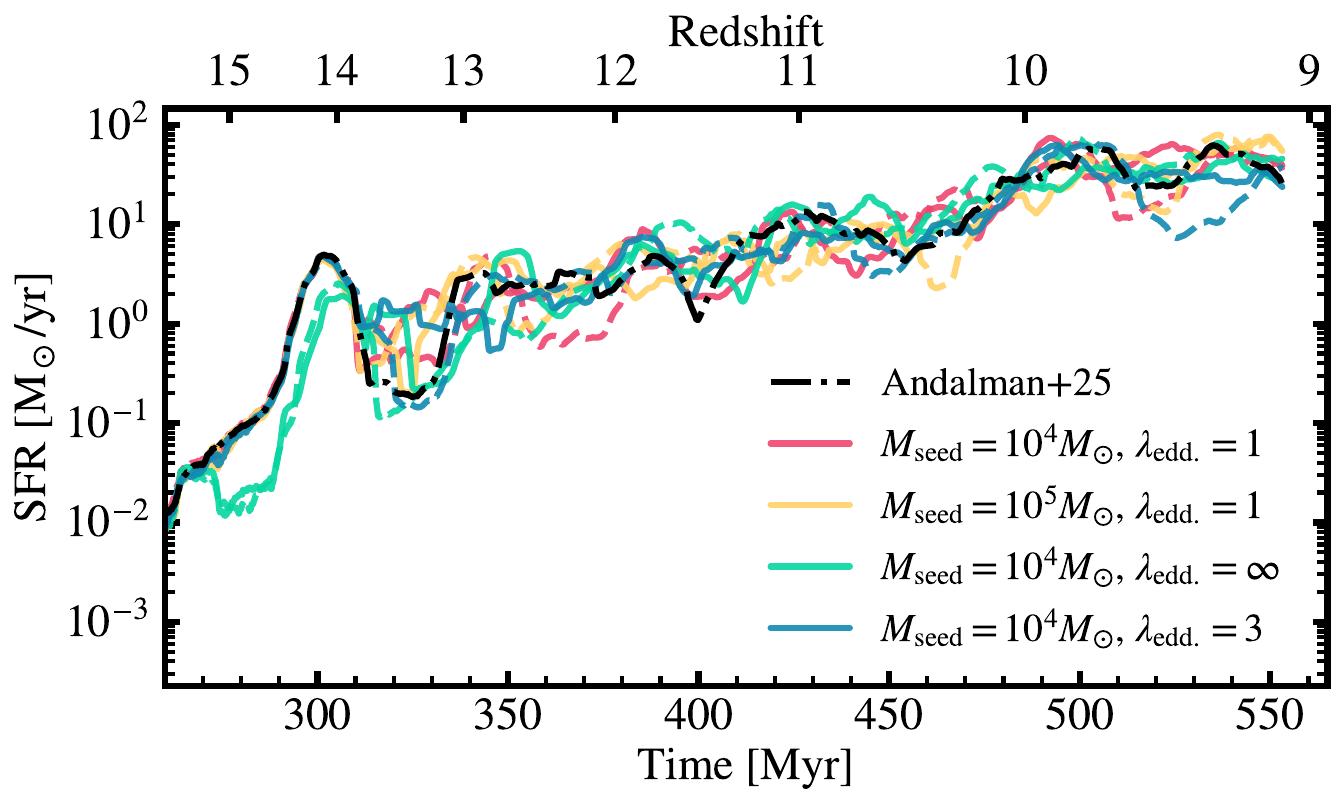}
    \caption{The SFR as a function of time over the duration of our simulations. This quantity is computed from stars within a 2 kpc box centred on the SMBH at the centre of the galaxy. All curves correspond to a running average with a window of 10 Myr. The black dash-dot curve is the measured SFR from the base fiducial simulation of \citet{andalman_origin_2025} with no SMBH. The solid coloured lines correspond to simulations with AGN feedback and the dashed coloured lines correspond to simulations without AGN feedback. In total there are SFRs for 9 simulations plotted in this figure as a function of time.}
    \label{fig:SFR_vs_time}
\end{figure}

During the calm accretion epoch, the environment of the SMBH is approximately equivalent for all simulations that have a \text{(super-)Eddington} limit. The characteristic density of this environment is $n_{\rm H} \sim 10^3 \; \rm cm^{-3}$ which implies $M^{\rm cool}_{\rm sink,crit} \sim 10^8 \; \rm M_\odot$. This implies the heating provided from AGN feedback is insignificant when compared to cooling in this epoch. Once stellar feedback turns on and the SMBH transitions into the bursty epoch of accretion, the sink sphere enters into a hot-diffuse pocket of the ISM and the density within the sink sphere decreases by $\sim 1-2$ dex which in turn lowers the critical cooling mass to $M^{\rm cool}_{\rm sink,crit} \sim 10^4 - 10^6 \; \rm M_\odot$ (see Fig. \ref{fig:mass_thresholds}). In these hot-diffuse patches it is then less clear how dominant or subdominant AGN heating is over cooling because the cooling mass is rapidly fluctuating above and below the SMBH mass by $\sim \pm 1$ dex. 

The subsequent $\sim 50 \; \rm Myrs$ are crucial in determining the fate of the SMBH because it is representative of what happens when the sink sphere enters into a hot-diffuse pocket of the ISM. During this period, the longer AGN heating is dominant over cooling ($M_{\rm sink} > M^{\rm cool}_{\rm sink,crit}$), the more capable the SMBH is of generating sufficient outward pressure to modify and mediate the ISM of the surrounding nuclear region. Mediation of the ISM occurs because AGN feedback sources thermal energy which drives pressure gradients that enhance mixing between the ISM phases in the nuclear region. This mediation has the effect of reducing the frequency and duration of abrupt increases in sink sphere density because it prevents the sink sphere from completely plunging into patches of the cold-dense ISM. The fraction of time that AGN heating dominates over cooling in this period is $\sim 30\%$ for the \texttt{edd\_AGN} simulation, $\sim 60\%$ for the \texttt{high\_seed\_edd\_AGN} simulation, $\sim 80\%$ for the \texttt{super\_edd\_AGN} simulation, and $100\%$ for the \texttt{no\_edd\_AGN} simulation. When we compare the sink sphere environment in simulations with and without AGN feedback (solid and dashed curves in Fig. \ref{fig:sink_sphere_env}) between $t \sim 300 \; \rm Myr$ and $t \sim 350 \; \rm Myr$, it is clear that the SMBH in the \texttt{no\_edd\_AGN} simulation is the most capable of altering the nuclear environment and the SMBH in the \texttt{edd\_AGN} simulation is the least capable of altering the nuclear environment through turbulent mixing. 

If the ISM in the nuclear environment is sufficiently pacified in the initial $\sim 50 \; \rm Myrs$ of the bursty accretion epoch, then there is enough time for the sound speed to rise as runaway heating occurs (see equation \ref{eq: sound_speed_vs_t}). Once the sound speed approaches the escape velocity of the halo the SMBH becomes capable of unbinding sink sphere gas from the halo, regulating its own growth to achieve self-regulation. The SMBH then grows in step with the self-regulated mass $M^{\rm esc}_{\rm sink,crit}$ as shown in Fig. \ref{fig:SMBH_mass_vs_time}. 

We attribute the difference in amplitude between the self-regulated mass and the mass of SMBHs that achieve self-regulation to an assumption of the \cite{biernacki_dynamics_2017} model. The original form of the model assumes that there is little stochasticity in the ISM environment and that cold-dense gas is distributed nearly isotropically around the SMBH. In our MDG environment, this assumption is not necessarily true. The ISM of the MDG is multiphase and turbulent which means that the sink sphere could be straddling between cold-dense patches and hot-diffuse pockets of the ISM at any given time within the bursty accretion epoch. We provide an qualitative example of this picture in the rightmost zoom-in panel of Fig. \ref{fig:SMBH_mass_vs_time}. 

The fraction of the sink sphere that is in a hot-diffuse pocket of the ISM will be capable of the runaway heating required for self-regulation while the remainder of the sink sphere is incapable of heating cold-dense gas. This leads to venting as AGN heated gas follows a path of least resistance through the hot-diffuse ISM. When compared to the isotropic assumption, this effect reduces the mass required to heat hot-diffuse gas to the escape velocity because there is less gas which needs to be heated. This in turn lowers the effective self-regulation mass $M^{\rm esc}_{\rm sink,crit}$ by a factor $0 \leq f_{\rm iso} \leq 1$ that accounts for the fraction of solid angles the AGN heated gas can escape the sink sphere through. Empirically, we find this fraction to be $f_{\rm iso} \sim 0.7$ in the calm accretion epoch ($t \lesssim 300 \; \rm Myr$) and $f_{\rm iso} \sim 0.5$ in the bursty accretion epoch ($t \gtrsim 300 \; \rm Myr$). 

While there is some variation in the growth of the SMBH across all the self-regulated simulations early on, by $\sim 380 \; \rm Myr$, they all follow the same evolutionary growth track determined by the self-regulated mass. If the growth of the SMBH is described by self-regulation, then the resulting mass can be tuned using the thermal energy coupling efficiency parameter $\epsilon_{\rm c}$. According to equation \ref{eq:Mesc}, the self-regulated mass has the proportionality $M^{\rm esc}_{\rm sink,crit} \propto r_{\rm sink} \epsilon_{\rm c}^{-1/2}$. Using this relation, if we want $M_{\rm sink} \sim 10^7 \; \rm M_\odot$ by $z \sim 9$ at fixed resolution, then we need to decrease $\epsilon_{\rm c}$ by a factor of 100. Similarly, if we increase our minimum resolution to $\Delta x_{\rm min} = 1 \; \rm pc$, then we would need to reduce $\epsilon_{\rm c}$ by a factor of 100 to keep the resulting SMBH mass at $\sim 10^6 \; \rm M_\odot$.

In the \texttt{edd\_AGN} simulation, self-regulation does not happen. Instead the SMBH fails to achieve self-regulation in the initial $\sim 50 \; \rm Myrs$ of the bursty accretion epoch because an insufficient amount of time ($\lesssim 50\%$) is spent above the critical cooling mass. Surprisingly, this remains true even as the SMBH grows more massive than $\sim 10^5 \; \rm M_\odot$: the same mass that all the other simulations achieve self-regulation. We explain this result with the chemical evolution of the galaxy. The metallicity evolution in the sink sphere grows exponentially, achieving an order of magnitude of growth from $t \sim 300 \; \rm Myr$ to $t \sim 550 \; \rm Myr$. Since $M^{\rm cool}_{\rm sink,crit} \propto Z$, we expect the critical cooling mass to also grow secularly by an order of magnitude over the simulation. Coincidentally, in the bursty accretion epoch the SMBH also accretes exponentially at a rate of $\dot{M}_{\rm acc}\approx \alpha \lambda_{\rm edd.} \dot{M}_{\rm edd.}$ which translates to an order of magnitude increase in mass over the same window of time when $\lambda_{\rm edd.} = 1$ and $\alpha = 0.5$ (fiducial ISM). In other words, at fixed gas density, the SMBH mass and the critical cooling mass grow together from $t \sim 300 \; \rm Myr$ to $t \sim 550 \; \rm Myr$. This implies that the SMBH in the \texttt{edd\_AGN} simulation is simply not capable of achieving self-regulation at Cosmic Dawn because the critical mass required to overcome cooling is a moving goal post that cannot be passed. We will discuss this result further in Section \ref{subsec:timescales}.

\subsubsection{Comparison to Observations} 
\label{subsubsec:comparison_to_obs}
We compare the simulated SMBH masses in our simulation suite to recent \textit{JWST} observations of AGN at Cosmic Dawn ($z > 9$) in Fig. \ref{fig:SMBH_mass_vs_time}. These observations yield mass estimates of 5 SMBHs with masses $\gtrsim 10^6 \; \rm M_\odot$: UHZ-1 \citep{goulding_uncover_2023, natarajan_first_2024}, GHZ9 \citep{napolitano_dual_2025, napolitano_seven_2025}, CAPERS-LRD-z9 \citep{taylor_capers-lrd-z9_2025}, MoM-z14 \citep{naidu_black_2025} and GN-z11.\footnote{There is an ongoing debate on the nature of GN-z11 based on different interpretations of the spectral energy distribution \citep{maiolino_small_2024, bunker_jades_2023}.} While not all of these measurements are accompanied by robust stellar mass estimates, the stellar mass estimates of these host galaxies are within $M_\star \sim 10^{8-9} \; \rm M_\odot$ which is comparable to our simulated MDG. 

Aside from GN-z11, our self-regulated SMBHs are 1-2 dex lower in mass than the \textit{JWST} SMBH mass estimates. At our resolution, we require coupling to be roughly $100\times$ less efficient $\epsilon_{\rm c} \lesssim 0.15 \%$ for the self-regulated mass $M^{\rm esc}_{\rm sink,crit}$ to be consistent with these observations (assuming $f_{\rm iso} \sim 0.5$). This would also increase $M^{\rm cool}_{\rm sink,crit}$ by the same factor. This implies that when $\epsilon_{\rm c} \lesssim 0.15 \%$, any Eddington limited SMBH ($\lambda_{\rm edd.} = 1$) with a plausible seed mass ($M_{\rm seed} \lesssim 10^6 \; \rm M_\odot$) would be incapable of achieving self-regulation because it would be incapable of generating sufficient heat to overtake cooling, preventing runaway heating. 

We note that we assume a fixed radiative efficiency $\epsilon_{\rm r} = 10\%$ and fixed resolution $r_{\rm sink} = 10 \; \rm pc$ when making this statement. If we were to increase the minimum resolution to $\Delta x_{\rm min} \sim 1 \; \rm pc$ then we would require a much smaller coupling efficiency to raise the self-regulated mass to be consistent with observations which in-turn could also prevent self-regulation. It is unclear if this would be true as increasing resolution can have other unforeseen effects (e.g. gas density is smoothed over the resolution scale). Alternatively, if we were to lower the radiative efficiency to $\epsilon_{\rm r} = 1\%$ then we would expect $\dot{M}_{\rm edd.}$ to increase by a factor of 10. In this scenario, any seed mass could achieve self-regulation while accreting at the Eddington rate. 

With these assumptions and caveats in mind, we identify only one clear possible evolutionary pathway in our simulations for the SMBH that results in masses comparable to MoM-z14, GN-z11, UHZ-1, GHZ9, and CAPERS-LRD-z9 while being self-regulated in the MDG environment. The pathway requires that the SMBH is seeded with $M_{\rm seed} \gtrsim 10^4 \; \rm M_\odot$ at $z \sim 15$, undergoes marginally super-Eddington accretion $\lambda_{\rm edd.} \gtrsim 3$, and heats the surrounding environment inefficiently $\epsilon_{\rm c} \lesssim 0.15 \%$. This picture is broadly consistent with the findings of \cite{lupi_sustained_2024}, which show that it is possible for a SMBH to undergo sustained bursts of super-Eddington accretion while having inefficient AGN feedback in the high redshift universe. We note that one should be wary of the resolution dependence (and the tunability via $\epsilon_{\rm c}$ and $\epsilon_{\rm r}$) of the critical masses that dictate self-regulation in our simulations. Future work is required to identify fundamental fixed physical scales that are not resolution dependent which describe how feedback energy is deposited into the environment. Identifying these physical scales will require inclusion of missing physics not included in our simulations. 

\subsection{Impact of SMBH on the Host Galaxy}
\label{subsec:impact}

As discussed in \cite{andalman_origin_2025} and in Section \ref{sec:simulations}, the galaxy that forms without a SMBH at the galactic centre is a thick disc partially supported by rotation and partially supported by turbulence. The disc is filled with gas spanning out to $R_{\rm gas} \sim 700 \; \rm pc$ and stars spanning out to $R_{\rm stars} \sim 400 \; \rm pc$. The baryonic content within the galaxy is predominantly in cold-dense clusters of diameter $\sim 100 \; \rm pc$. The halo is growing nearly exponentially up to $z \sim 9$ due to a continuous supply of cold gas and dark matter from surrounding cosmic filaments. The continuous supply of cold gas allows for intense sustained star formation within dense stellar clusters inside the galaxy from $z \sim 15$ to $z \sim 9$. This description remains unchanged across all of our simulations. 

In the low-redshift Universe, AGN feedback quenches star formation in galaxies \citep{silk_quasars_1998, di_matteo_energy_2005, mountrichas_link_2023, goubert_role_2024, bluck_galaxy_2024}. One would expect that we would see substantial suppression in the SFR across the galaxy over time due to additional thermal energy being deposited into the galaxy from the SMBH but this is not what we find in our simulations. Instead, in Fig. \ref{fig:SFR_vs_time}, we find that the secular evolution of the SFR is left intact across all simulations. This is a departure from expectations, we provide further explanation for this behaviour in Section \ref{subsec:timescales} using relevant growth timescales in the simulated system. There are fluctuations in the SFR with a characteristic timescale of $\Delta t \lesssim 20 \; \rm Myr$ but these fluctuations do not suggest any systemic quenching of the galaxy.

In the simulations with no Eddington limit ({\texttt{no\_edd\_AGN}} and {\texttt{no\_edd\_no\_AGN}}) there is a substantial dip in star formation within the galaxy at $t \sim 300 \; \rm Myr$. When AGN feedback is turned on and there is no Eddington limit (solid green), the SMBH reaches self-regulation within $\Delta t \sim 1 \; \rm Myr$ and grows to $M_{\rm sink}  \simeq M^{\rm esc}_{\rm sink,crit} \sim 10^5 \; \rm M_\odot$. Once self-regulation is achieved, the SMBH injects thermal energy into the surrounding interstellar medium. At $t \sim 300 \; \rm Myr$ the galaxy is significantly smaller than at $t \sim 550 \; \rm Myr$ which allows AGN feedback to temporarily quench star formation in the few constituent stellar clusters close to the SMBH which constitute a significant fraction of the total stellar mass at that time. When AGN feedback is turned off and there is no Eddington limit (dashed green) we also see comparable quenching within the galaxy around $t \sim 300 \; \rm Myr$ in Fig. \ref{fig:SFR_vs_time}. In this case, the SMBH continues to grow at a rapid effective rate $\lambda_{\rm edd.} \sim 10^3 - 10^4$ reaching $\gtrsim 10^{7.5} \; \rm M_\odot$ by $t \sim 300 \; \rm Myr$, achieving the same effect as before by instead accreting all the cold gas close to the few dense stellar clusters rather than heating the gas via feedback. 

\begin{figure}
    \centering
    \includegraphics[width=\linewidth]{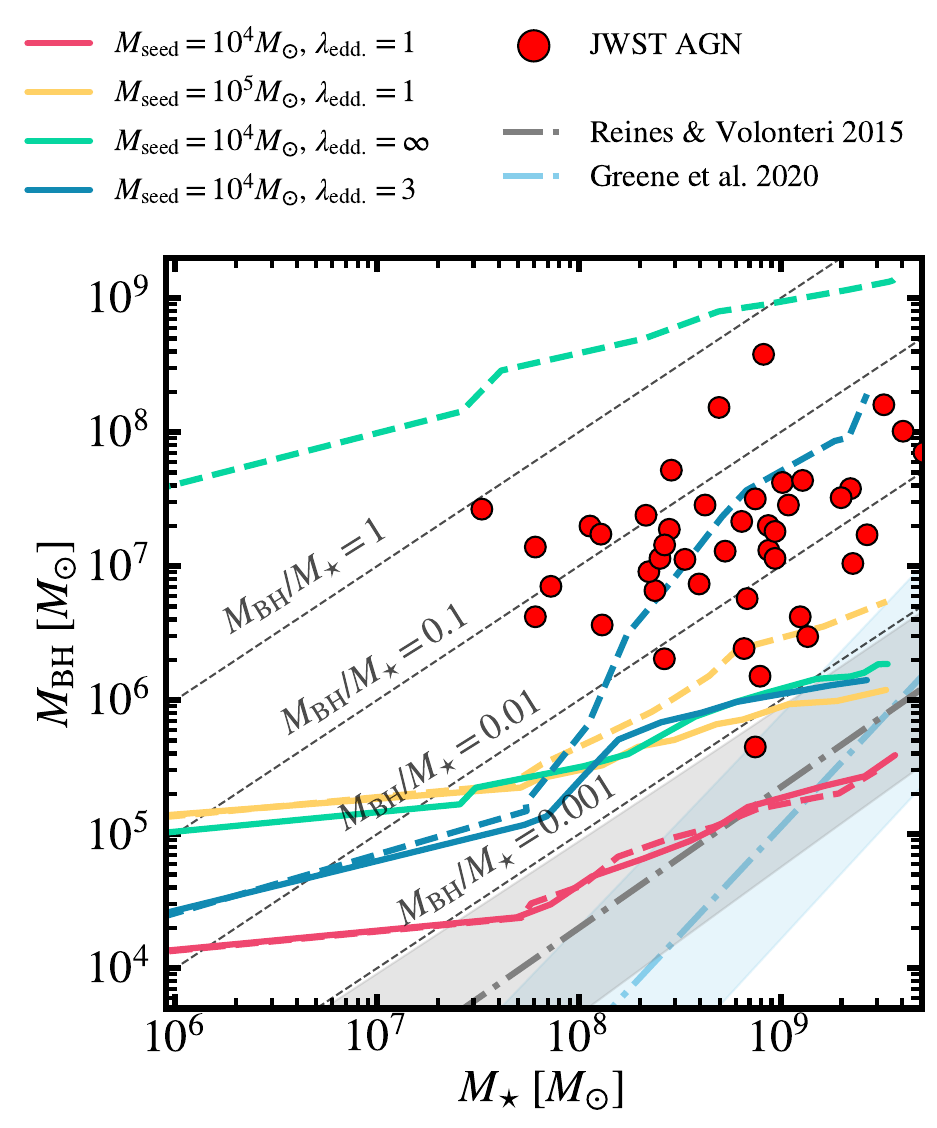}
    \caption{This figure shows the mass of the SMBH at the centre of our simulated MDG relative to the stellar mass enclosed within a sphere around the galaxy of radius 1 kpc. Solid lines denote simulations with AGN feedback turned on with $(\epsilon_{\rm c} = 15\%)$ and dashed lines indicate simulations with AGN feedback turned off $(\epsilon_{\rm c} = 0\%)$. In total we show how 8 of our simulations compare to observations of AGN between $4 < z < 11$ observed with \textit{JWST} \citep{ubler_ga-nifs_2023, harikane_jwstnirspec_2023, kokorev_uncover_2023, maiolino_jades_2024, juodzbalis_jades_2025}. These simulations vary the Eddington accretion limit by a factor of $\lambda_{\rm edd.}$ and the SMBH seed mass $M_{\rm seed}$. The green lines correspond to simulations where the Eddington limit was removed entirely. The dash-dot curves show local Universe relations from \citet{reines_relations_2015} and \citet{greene_2020_review}.}
    \label{fig:Mbh_vs_Mstar}
\end{figure}

The resilience of star formation within the galaxy can also be seen in the $M_{\rm BH}$-$M_{\star}$ relation. In Fig. \ref{fig:Mbh_vs_Mstar} we show the simulated trajectories on the $M_{\rm BH}$-$M_{\star}$ plane. Regardless of the growth of the SMBH across all simulations, the final stellar mass of the galaxy and its evolution remain unchanged relative to the fiducial simulation of \cite{andalman_origin_2025} (see their Figure 9). Stellar feedback becomes substantial once the galaxy reaches a stellar mass of $\sim 10^8 \; \rm M_\odot$ which occurs at $z \sim 14$ in our simulations. Even in the most extreme SMBH growth scenario ({\texttt{no\_edd\_no\_AGN}}, green dashed curve) there is no significant impact on the resulting stellar mass. The SMBH becomes $\sim 40\times$ larger than the stellar mass of the galaxy within 1 Myr after being seeded. Growth then slows: first from accreting all the gas within the sink sphere during the calm accretion epoch, then from stellar feedback regulating accretion after entering into the bursty accretion epoch. During this period of slowed SMBH growth, the galaxy continues to grow exponentially 
via accretion from cosmic filaments. The cold gas accreted from cosmic filaments fuels star formation in the outer regions of the galaxy, ultimately leaving no impact on the resulting stellar mass. 

Between simulations there are significant deviations in growth trajectories in the $M_{\rm BH}$-$M_{\star}$ plane. As a consequence of our seeding prescription described in Section \ref{subsec: SMBH formation}, all growth trajectories in Fig. \ref{fig:Mbh_vs_Mstar} start at $M_{\rm BH} / M_\star \sim 0.01 - 0.1$ \citep[high relative to local relations][]{reines_relations_2015, greene_2020_review}. The SMBH-galaxy co-evolution in our simulations intersects local relations by $z \sim 9$ when $\lambda_{\rm edd.} = 1$ or when self-regulation via AGN feedback is achieved. While these simulations do intersect local relations, they fall below the known population of high redshift AGN seen with \textit{JWST} \citep{ubler_ga-nifs_2023, harikane_jwstnirspec_2023, kokorev_uncover_2023, maiolino_jades_2024, juodzbalis_jades_2025} between $4 < z < 11$. In these simulations $\dot{M}_{\rm sink} \lesssim \dot{M}_{\star}$ for most of the simulation. Meanwhile, simulations with super-Eddington accretion and no AGN feedback can achieve $\dot{M}_{\rm sink} \gtrsim \dot{M}_{\star}$, allowing an upward climb in the $M_{\rm BH}$-$M_{\star}$ plane. The simulation that best matches the population of high redshift AGN seen by \textit{JWST} is {\texttt{super\_edd\_no\_AGN}}. We note that a simulation with AGN feedback can achieve a similar outcome so long as the self-regulated mass $M^{\rm esc}_{\rm sink, crit}$ is tuned via the thermal energy coupling efficiency parameter $\epsilon_{\rm c} \lesssim 0.15 \%$ and the SMBH is capable of super-Eddington accretion to grow in step with the self-regulated mass as discussed in Section \ref{subsec:SMBH Growth}.

\subsection{Insights From Timescales}
\label{subsec:timescales}

There are 3 relevant exponential growth timescales which we use to gain insights about our simulated system: the growth timescale of the SMBH $\tau_{\rm sal} \simeq 45 \; \mathrm{Myr} / \alpha \lambda_{\rm edd.}$, the growth timescale of the halo $\tau_{\rm halo}$, and the growth timescale of the self-regulation mass of the SMBH $\tau_{\rm esc}$. 

\begin{figure}
    \centering
    \includegraphics[width=\linewidth]{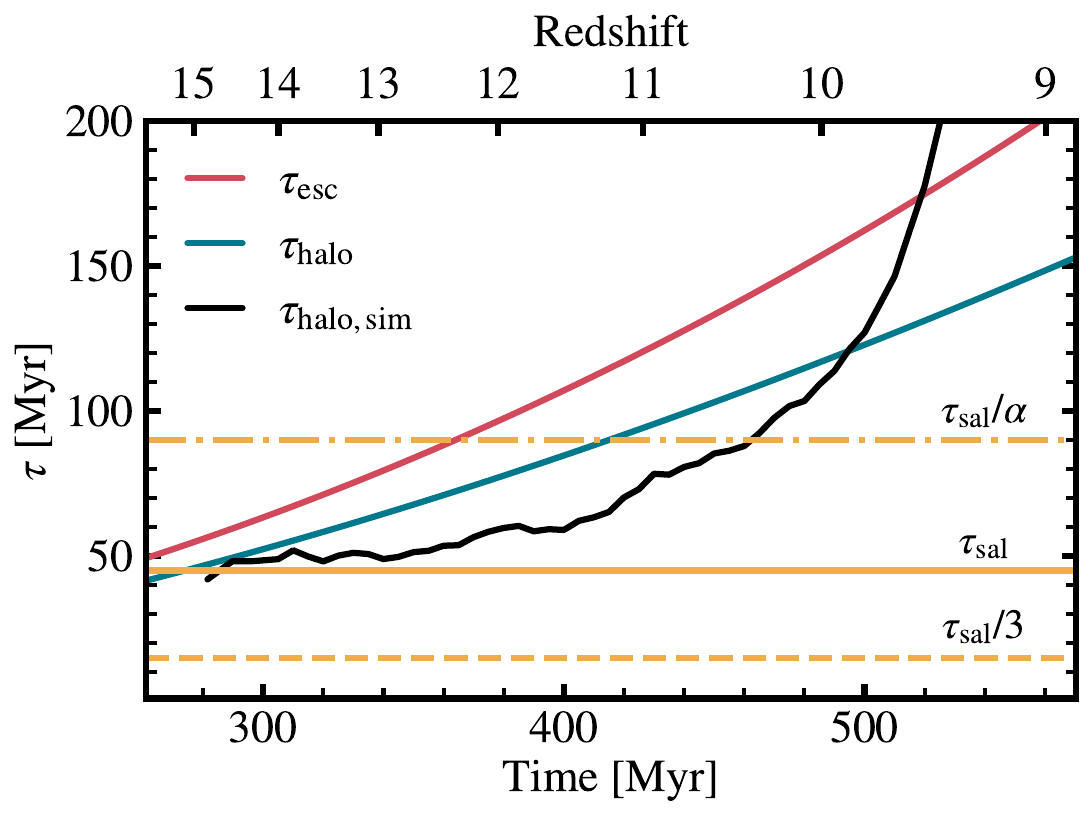}
    \caption{We show the exponential growth timescale $\tau$ as a function of time (or redshift) for the critical self-regulation timescale $\tau_{\rm esc}$ (red), the halo $\tau_{\rm halo}$ (blue), and the Eddington limit $\tau_{\rm sal}$ (yellow). The growth of the halo measured in the simulation $\tau_{\rm halo, sim}$ is depicted in black and can be approximated by equation \ref{eq:Mdot_halo_timescale} at early times. We note that as time progresses the growth of the halo slows down while the Salpeter timescale remains constant.}
    \label{fig:timescales}
\end{figure}

According to \citep{dekel_toy_2013, correa_accretion_2015, andalman_origin_2025}, at Cosmic Dawn, a halo of mass $M_{\rm halo}$ will accrete exponentially with redshift at a specific accretion rate 
\begin{equation}
    \frac{\dot{M}_{\rm halo}}{M_{\rm halo}} \simeq s \left( \frac{M_{\rm halo}}{10^{12} \; \rm M_\odot} \right)^\beta (1 + z)^{5/2}  = \frac{1}{\tau_{\rm halo}}\; .
\label{eq:Mdot_halo_timescale}
\end{equation}
We define $s$ as the specific accretion rate onto a $M_{\rm halo} = 10^{12} \rm M_\odot$ halo at $z = 0$ and $\beta$ to be the power law of the fluctuation power spectrum. These parameters are found to be $s \simeq 0.03 \; \rm Gyr^{-1}$ and $\beta \simeq 0.14$ in \cite{dekel_toy_2013}. As time progresses the exponential growth of the halo slows down $\tau_{\rm halo} \propto (1 + z)^{-5/2}$ which can be seen in Fig. \ref{fig:timescales}. We measure the growth timescale of the halo in the fiducial simulation of \cite{andalman_origin_2025} with no additional SMBH influence (black curve) and find the analytic form is a valid but rough approximation to our simulations.

The halo growth timescale is crucial in describing the evolution of our critical mass thresholds ($M^{\rm cool}_{\rm sink, crit}$ and $M^{\rm esc}_{\rm sink,crit}$) derived in Section \ref{subsec:regimes}. As mentioned in Section \ref{subsec:regimes} and \ref{subsec:SMBH Growth}, the critical cooling-to-heating mass threshold grows with metallicity. Empirically we find the metallicity of the gas within the galaxy (and the sink sphere) is well described by exponential growth with growth timescale $\tau_{\rm halo}$. One would expect this to be the case because metal enrichment is a direct result from the recycling of gas into stars which in-turn depends on the fresh supply of cold gas onto the halo. In other words, we would expect $\dot{Z} \propto \dot{M}_\star \propto \dot{M}_{\rm halo}$. Therefore, we expect that the critical cooling mass threshold evolves with time proportionally to the halo
\begin{equation}
    \left.\frac{\dot{M}^{\rm cool}_{\rm sink,crit}}{M^{\rm cool}_{\rm sink,crit}} \right\vert_{n_{\rm H}} = \frac{\dot{Z}}{Z} \approx \frac{1}{\tau_{\rm halo}} \; .
\end{equation}
We note that this is a simplistic approximation that ignores outflows from stellar feedback. These outflows can induce fluctuations in the metallicity on timescales that are short relative to $\tau_{\rm halo}$. We ignore these fluctuations as they are not important in describing the secular evolution of the critical cooling mass threshold. 

We can express the growth of the self-regulated mass using the growth of the halo. With a little manipulation, we can rewrite equation \ref{eq:Mesc} assuming that the Universe at Cosmic Dawn is well described by an Einstein-deSitter universe such that $\bar{\rho}_{\rm m} = \bar{\rho}_{\rm c} = 3 H^2(z) / 8 \pi G$ and $a \propto t^{3/2}$. When we do so, the self-regulated mass can be written as
\begin{equation}
    M^{\rm esc}_{\rm sink,crit} = \frac{2 H(z) M_{\rm halo} r_{\rm sink}}{c} \left[ \frac{2 \Delta}{15 \epsilon_{\rm r} \epsilon_{\rm c}}\right]^{1/2} \propto H(z) M_{\rm halo} \; .
\label{eq: Mesc_simple}
\end{equation}
In this form it is clear that the self-regulated mass is proportional to the mass of the halo $M_{\rm halo}$ and the Hubble parameter $H(z) \simeq H_0 \sqrt{\Omega_{\rm m,0}} (1 + z)^{3/2}$. This implies that $M^{\rm esc}_{\rm sink, crit}$ grows exponentially with timescale
\begin{equation}
    \frac{1}{\tau_{\rm esc}} = \frac{\dot{M}^{\rm esc}_{\rm sink, crit}}{M^{\rm esc}_{\rm sink, crit}}  = \frac{1}{\tau_{\rm halo}} - \frac{3}{2} H(z)\; .
\label{eq:tau_escape}
\end{equation}
We note that, although the values of $M^{\rm cool}_{\rm sink,crit}$ and $M_{\rm sink, crit}^{\rm esc}$ are explicitly resolution dependent, their growth timescales ($\tau_{\rm halo}$ and $\tau_{\rm esc}$) are not resolution dependent quantities.

We show these 3 growth timescales as a function of time in Fig. \ref{fig:timescales}. The Salpeter timescale $\tau_{\rm sal}$ remains constant while $\tau_{\rm esc}$ and $\tau_{\rm halo}$ grow with time, implying a decrease in the rate of growth. Coincidentally, $\tau_{\rm halo}$ and $\tau_{\rm sal}$ are comparable until the transition from the calm accretion epoch to the bursty accretion epoch ($t \sim 300 \; \rm Myr$). The measured halo growth timescale in our simulation $\tau_{\rm halo, sim}$ remains comparable to $\tau_{\rm sal}$ until $t \sim 350 \; \rm Myr$. In the bursty accretion epoch the turbulent-multiphase ISM induces starvation modes which increases the effective growth timescale by a factor of $1/\alpha$. This extends the time spent in the simulation where $\tau_{\rm halo} \lesssim \tau_{\rm sal}/\alpha$ to $t \sim 460 \; \rm Myr$ for the simulation ($\tau_{\rm halo,sim}$, black curve) and $t \sim 420 \; \rm Myr$ for the analytical form ($\tau_{\rm halo}$, blue curve).

Since $M_{\rm halo} \gg M_{\rm sink}$, a comparable growth timescale implies that $\dot{M}_{\rm halo} \gg \dot{M}_{\rm sink}$. If we assume that $M_{\star} = \epsilon_{\rm int} f_{\rm b} M_{\rm halo}$, where $\epsilon_{\rm int} \gtrsim 10 \%$ is the integrated SFE (see \cite{andalman_origin_2025}) and $f_{\rm b} = \Omega_{\rm b,0} / \Omega_{\rm m,0}$ is the universal baryon fraction, then we expect $\dot{M}_{\star} \gg \dot{M}_{\rm sink}$. By writing the ratio explicitly
\begin{equation}
    \frac{\dot{M}_\star}{\dot{M}_{\rm sink}} \approx \frac{M_\star}{M_{\rm sink}} \left(\frac{\tau_{\rm sal} /\alpha \lambda_{\rm edd.}}{\tau_{\rm halo}} \right) \; ,
\end{equation}
we can see that it is $\gtrsim 10^3 \gg 1$ for the entire duration of our simulation (assuming $\lambda_{\rm edd.} = 1$ and $\alpha = 0.5$). This is confirmed by the rightward SMBH-galaxy co-evolution trajectories in Fig. \ref{fig:Mbh_vs_Mstar}. Only when $\lambda_{\rm edd.} > 1$ can $\dot{M}_{\rm sink} > \dot{M}_\star$ during the bursty accretion epoch in our MDG environment. 

The halo is capable of accreting dark matter and cold gas from cosmic filaments at a rate that is fast enough to continuously replenish any gas in the galaxy that might be accreted onto the SMBH, even at a maximal accretion rate (unless $\lambda_{\rm edd.} > 1$). Furthermore, any heating done by AGN feedback is not capable of quenching star formation across the galaxy because there is a constant fresh supply of cold gas to replace any pre-existing heated gas. In other words, AGN feedback cannot regulate star formation on a galactic scale unless the growth timescale of the halo is much larger than the maximal growth timescale of the SMBH ($\tau_{\rm halo} \gg \tau_{\rm sal}/\alpha\lambda_{\rm edd.}$). When this is true, AGN feedback can regulate star formation because the gas used to form stars in the galaxy cannot be immediately replaced by cold gas accreted onto the galaxy. This provides a fundamental explanation for why, across all of our simulations, there is no secular quenching of the SFR in Fig. \ref{fig:SFR_vs_time}. 

This behaviour will not persist indefinitely. As time progresses, $\tau_{\rm halo}$ grows with time, therefore $\dot{M}_{\star}$ decreases with time. Eventually, halo growth will slow to a rate that is unsustainable to replenish the cold gas reservoir needed for a sustained SFR of $10-100 \; \rm M_\odot/yr$. As a result, star formation will slow down, stellar feedback will heat up the ISM, and the intense turbulent-multiphase ISM seen in our simulations will relax. As the ISM begins to stabilize we expect $f_{\rm iso}$ and $\alpha$ to slowly increase back to unity. As this happens, the SMBH-galaxy co-evolution will be better described by the standard model of \cite{biernacki_dynamics_2017}. We also expect that rightward movement in Fig. \ref{fig:Mbh_vs_Mstar} will slow as the SFR drops. As this happens, self-regulated SMBHs will slowly crawl up the local Universe relations \citep{reines_relations_2015, greene_2020_review} in the $M_{\rm BH}-M_\star$ plane. 

In simulations that do not enter the self-regulated regime ({\texttt{edd\_AGN}} and {\texttt{edd\_no\_AGN}}) at Cosmic Dawn, the picture is different. As discussed in Section \ref{subsec:SMBH Growth}, stellar feedback drives the properties of the ISM which regulates SMBH growth after entering into the bursty accretion epoch, not AGN feedback. This effect is captured by multiplying the Eddington rate by a factor $\alpha \approx 0.5$ (see Fig. \ref{fig:SMBH_mass_vs_time}). We show in Fig. \ref{fig:timescales} that coincidentally, the effective growth timescale of the SMBH $\tau_{\rm sal}/\alpha \lambda_{\rm edd.}$ is greater than or comparable to the halo growth time scale for most of our simulation (when $\lambda_{\rm edd.} = 1$ and $\alpha = 0.5$). We expect that when $\tau_{\rm halo} \gg \tau_{\rm sal} / \alpha$, the mass of the SMBH will be able to grow faster than $M^{\rm cool}_{\rm sink,crit}$ and eventually catch up by $z \sim 6-7$ when $M_{\rm sink} \sim 10^7 \; \rm M_\odot$. Once this happens then the SMBH will achieve runaway heating and begin self-regulation. 

Since $\tau_{\rm esc}$ increases with time faster than $\tau_{\rm halo}$, we expect the self-regulated mass to grow slowly relative to the critical cooling mass threshold. By the time $M_{\rm sink} \gtrsim M^{\rm cool}_{\rm sink,crit}$, it is entirely possible that $M^{\rm cool}_{\rm sink,crit}$ could be greater than the self-regulation mass $M^{\rm esc}_{\rm sink,crit}$. If this is true, then the SMBH would need to become more massive than $M^{\rm esc}_{\rm sink,crit}$ to become self-regulated. This implies that a SMBH that is not capable of achieving self-regulation at Cosmic Dawn could grow to be larger than a SMBH that did achieve self-regulation early on. If the thermal energy coupling efficiency was lowered $\epsilon_{\rm c} \lesssim 0.15 \%$ then the result could be more extreme. The self-regulated mass would be raised by a factor of 10 while the critical cooling mass threshold would be raised by a factor of 100. This would increase the amount of time needed for $M_{\rm sink} \gtrsim M^{\rm  cool}_{\rm sink,crit}$, allowing the resulting SMBH to be even more massive than SMBHs that achieved self-regulation at Cosmic Dawn. The SMBH overgrowth that results from this process creates a violent burst of feedback that halts SMBH growth until $M_{\rm sink} \sim M^{\rm esc}_{\rm sink,crit}$. From this example we learn that, if the SMBH does not become massive enough for AGN feedback to be a dominant growth regulator before star formation turns on in the galaxy, then the SMBH will have to wait until after Cosmic Dawn to become self-regulated and it might become more massive than a SMBH that became self-regulated at Cosmic Dawn. 

\section{Conclusions}
\label{sec:conclusions}
In this work we conduct a suite of numerical experiments to explore how SMBHs grow in and influence massive galaxies ($M_\star \gtrsim 10^8 M_\odot$) at Cosmic Dawn ($z \sim 15 - 9$). The environment which this study resides is a large overdensity which forms a $\sim 10^{11} M_\odot$ halo by $z \sim 9$. Our suite contains 11 simulations varying the strength of AGN feedback ($\epsilon_{\rm c} = 0 \%$ or $\epsilon_{\rm c} = 15 \%$), the seed mass ($M_{\rm seed} = 10^4 \; \rm M_\odot$ or $M_{\rm seed} = 10^5 \; \rm M_\odot$), maximal accretion rate ($\lambda_{\rm edd.} = 1, 3, \infty$), and the strength of stellar feedback. By default our simulations use the fiducial setup of \cite{andalman_origin_2025} as our base MDG (see Fig. \ref{fig:zoom-in}) which the SMBH resides in and interacts with. Importantly, this base MDG model does not assume a fixed star formation rate, but instead uses a physically-motivated turbulence-based subgrid multi-freefall model of star formation. This model enables variable local star formation efficiencies which can reach as high as $\lesssim 20 \%$, leading to more stars being created within the galaxy than would be achieved assuming a constant star formation efficiency commonly found in the local Universe. Our key findings from this work are listed below. 

\begin{itemize}
    \item \textbf{SMBH Growth in the MDG Environment:} Once star formation turns on in the galaxy, the ISM in the nuclear region becomes turbulent and multiphase due to stellar feedback. We denote this epoch as the bursty epoch of accretion ($t \gtrsim 300 \; \rm Myr$). The ISM in the bursty epoch of accretion consists primarily of patches of cold-dense and hot-diffuse gas which allow the SMBH to feast or starve as the sink sphere passes through these patches respectively (see Figures \ref{fig:SMBH_mass_vs_time} and \ref{fig:sink_sphere_env}). The stochasticity of these feast and starvation accretion modes is determined by the stochasticity of the SFR and the strength of stellar feedback. In the fiducial MDG environment we find that the SMBH spends approximately equal time in the feast mode as the starvation mode during the bursty epoch of accretion ($\alpha \approx 0.5$). We provide a simple analytical treatment to account for the role of the turbulent-multiphase MDG environment in SMBH growth (assuming AGN feedback to be weakly coupled to the gas or non-existent) by multiplying the Eddington rate by $\alpha$. 
    \newline
    \item \textbf{To Self-Regulate or Not to Self-Regulate:} It is not guaranteed that the SMBH will achieve self-regulation via AGN feedback in the MDG environment. We find that the SMBH must become massive enough (either by accretion or by seeding) for AGN heating to dominate over cooling within the sink sphere ($M_{\rm sink} \gtrsim M^{\rm cool}_{\rm sink, crit}$) while in the hot-diffuse patches of the ISM. The capability of the SMBH to modify and mediate the nuclear region of the ISM via AGN feedback is proportional to the amount of time spent above the critical cooling mass threshold (see Figures \ref{fig:mass_thresholds} and \ref{fig:sink_sphere_env}). If $\gtrsim 50\%$ of the time in hot-diffuse pockets is spent above the critical cooling mass threshold then runaway heating occurs within the sink sphere. As the hot-diffuse gas is super-heated, the gas is anisotropically ejected out of the halo. The mass evolution of the SMBH is then well described by the self-regulated mass $M^{\rm esc}_{\rm sink,crit}$. If the SMBH did not become massive enough to achieve self-regulation at Cosmic Dawn it will be forced to wait until a later time to do so. This evolutionary pathway could end in a higher resulting SMBH mass than SMBHs that achieve self-regulation at Cosmic Dawn.
    \newline
    \item \textbf{Impact on the Host Galaxy:} In our simulations AGN feedback has no effect on the secular evolution of the SFR (see Fig. \ref{fig:SFR_vs_time}) and the resulting stellar mass of the galaxy by $z \sim 9$ (see Fig. \ref{fig:Mbh_vs_Mstar}). This is a departure from expectations that we explain through comparing exponential growth timescales (see Fig. \ref{fig:timescales}). The SMBH cannot regulate star formation in the galaxy because the exponential growth rate of the halo (which sets the SFR, and thus the strength of stellar feedback) is sufficiently high that any heat transferred to the surrounding gas by AGN feedback is futile against the constant replenishing stream of fresh cold gas from surrounding cosmic filaments via halo accretion.
    \newline 
    \item \textbf{Comparisons to Observations:} We compare our simulations to observations of high redshift AGN seen by \textit{JWST}. Of all the simulations in our suite, we find the {\texttt{super\_edd\_no\_AGN}} simulation ($M_{\rm seed} \sim 10^4 \; \rm M_\odot$, $\lambda_{\rm edd.} = 3$) to best match the observed populations of AGN seed by \textit{JWST} (see Figures \ref{fig:SMBH_mass_vs_time} and \ref{fig:Mbh_vs_Mstar}). At our resolution, we determine that a low thermal energy coupling coefficient $\epsilon_{\rm c} \lesssim 0.15 \%$ is necessary for a self-regulated SMBH to become comparable to the mass estimates of MoM-z14, GN-z11, UHZ-1, GHZ9, and CAPERS-LRD-z9. Additionally, our numerical experiments indicate that super-Eddington accretion ($\lambda_{\rm edd.} \gtrsim 3$) is necessary for the SMBH to become massive enough to enter the self-regulated regime if $\epsilon_{\rm c} \lesssim 0.15 \%$ without requiring a seed mass above predictions from direct collapse scenarios ($M_{\rm seed} \gtrsim 10^6 \; \rm M_\odot$).
\end{itemize}

Going forward, pushing to a higher spatial resolution $\Delta x_{\rm min} \sim 1 \; \rm pc$ would be a valuable test to validate our theory and our findings as some of our results are necessarily resolution dependent. There are several additional physical effects which remain to be modelled in these simulations which could all impact our results. The inclusion of a more sophisticated accretion-dependent feedback prescription including both kinetic and radiative feedback with \textsc{ramses-rt} will be an important next-step in our simulations which could affect the strength of AGN feedback \citep{massonneau_how_2023,lupi_sustained_2024, husko_effects_2025, sanati_rapid_2025} which in turn could alter the SMBH growth behaviour significantly. Radiative feedback from stars could also radically change the structure of the ISM and thus SMBH-galaxy co-evolution \cite{ferrara_2025_noffb}. Additionally, our work assumes equilibrium thermochemistry between hydrogen, helium, and metals but accounting for non-equilibrium thermochemistry with \textsc{ramses-rt} can have a significant impact on our results \citep{rosdahl_ramses-rt_2013}. Ultimately, the inclusion of higher spatial resolution, more sophisticated feedback models, radiative transfer, non-equilibrium thermochemistry, would bring us closer to making direct predictions of the most massive galaxies and AGN at Cosmic Dawn.

\section*{Acknowledgements}
We would like to thank Sasha Tchekhovskoy, Helena Treiber, and Ronan Hix for insightful discussions. This material is based upon work supported by the National Science Foundation (NSF) and the U.S.-Israel Binational Science Foundation (BSF) under Award Number 2406558 and Award Title "The Origin of the Excess of Bright Galaxies at Cosmic Dawn". We are pleased to acknowledge that the work reported on in this paper was substantially performed using the Princeton Research Computing resources at Princeton University which is consortium of groups led by the Princeton Institute for Computational Science and Engineering (PICSciE) and Research Computing. Lastly, JS acknowledges and is thankful for the support from the Fannie \& John Hertz Foundation. 

\section*{Data Availability}
Each simulation generates at least 1 TB of output data. The total required storage is too large to host our data online. Data will be made available upon reasonable request to the corresponding author. 



\bibliographystyle{mnras}
\bibliography{biblio} 

\bsp	
\label{lastpage}
\end{document}